\documentclass[iop]{emulateapj}
\usepackage{apjfonts}
\usepackage{natbib}

\usepackage{graphicx}
\usepackage[usenames,dvips]{color}
\usepackage{floatrow}

\newcommand{\Msun}{\ifmmode {M_{\odot}}\else${M_{\odot}}$\fi}
\newcommand{\Lsun}{\ifmmode {L_{\odot}}\else${L_{\odot}}$\fi}
\newcommand{\Rsun}{\ifmmode {R_{\odot}}\else${R_{\odot}}$\fi}
\newcommand{\dt}{\ifmmode {\Delta \text{Log}T_{\text{eff}}}\else{$\Delta$Log$T_{\text{eff}}$}\fi}
\newcommand{\lt}{\ifmmode {\text{Log}T_{\text{eff}}}\else{Log$T_{\text{eff}}$}\fi}

\def\Chandra{${\it Chandra}$}

\shorttitle{Surface temperature inhomogeneities}
\shortauthors{Elshamouty {\it et al.}}

\begin{document}
\title{The impact of surface temperature inhomogeneities on quiescent neutron star radius measurements}

\author{%
  K.~G. Elshamouty\altaffilmark{1}, C.~O. Heinke\altaffilmark{1}, S.~M. Morsink\altaffilmark{1}, S. Bogdanov\altaffilmark{2}, A.~L. Stevens\altaffilmark{1,3},
}

\altaffiltext{1}{Department of Physics, University of Alberta, CCIS 4-181, Edmonton, AB T6G 2E1, Canada; alshamou@ualberta.ca}
\altaffiltext{2}{Columbia Astrophysics Laboratory, Columbia University, 550 West 120th Street, New York, NY 10027, USA}
\altaffiltext{3}{Anton Pannekoek Institute, University of Amsterdam, Postbus 94249, 1090 GE Amsterdam, the Netherlands}

% ================================= // BEGING ABSTRACT // =================================

\begin{abstract}

Fitting the thermal X-ray spectra of neutron stars (NSs) in quiescent X-ray binaries can constrain the masses and radii of NSs.
The effect of undetected hot spots on the spectrum, and thus on the inferred NS mass and radius, has not yet been explored for appropriate atmospheres and spectra.  A hot spot would harden the observed spectrum, so that spectral modeling tends to infer radii that are too small.  However, a hot spot may also produce detectable pulsations.
 
We simulated the effects of a hot spot on the pulsed fraction and spectrum of the quiescent NSs X5 and X7 in the globular cluster 47 Tucanae, using appropriate spectra and beaming for hydrogen atmosphere models, incorporating special and general relativistic effects, and sampling a range of system angles. 
We searched for pulsations in archival Chandra HRC-S observations of X5 and X7, placing 90\% confidence upper limits on their pulsed fractions below 16\%.
We use these pulsation limits to constrain the temperature differential of any hot spots, and to then constrain the effects of possible hot spots on the X-ray spectrum and the inferred radius from spectral fitting.  We find that hot spots below our pulsation limit could bias the spectroscopically inferred radius downward by up to 28\%. 
For Cen X-4 (which has deeper published pulsation searches), an undetected hot spot could bias its inferred radius downward by up to 10\%. 
Improving constraints on pulsations from quiescent LMXBs may be essential for progress in constraining their radii.
\end{abstract}

\maketitle
% ================================= // END ABSTRACT // =================================

%%%%%%%%%%%%%%%   INTRODUCTION   %%%%%%%%%%%%%%%

\section{Introduction}\label{s:intro}

One of the most intriguing unsolved questions in physics is the equation of state (EOS) of cold,  supranuclear-density matter which lies in the cores of neutron stars (NSs). Since each proposed EOS allows a limited range of values for the NS mass $M$ and radius $R$, accurate measurements of 
$M$ and $R$ can be used to constrain the NS EOS \citep[see for reviews:][]{Lattimer07, Hebeler13,Ozel13,Lattimer16,Haensel16,Steiner16}.

While it is possible, in some cases, to obtain accurate NS mass measurements \citep[e.g.][]{Demorest10,Freire11,Antoniadis13,Ransom14}, it is difficult to determine the NS radius. One method for determining the size of a NS is through a modification of the blackbody radius method \citep{Paradijs79}: if the distance, flux and temperature of a perfect blackbody sphere can be measured, then its radius is also known. Since NSs are not perfect blackbodies, this method has been modified to take into account more realistic spectra.  General relativistic effects also make these radius measurements degenerate with mass, providing constraints along curved tracks close to lines of constant $R_{\infty}=R/\sqrt{1-2GM/(Rc^2)}$, where $M$ and $R$ are the NS mass and radius. 

The two main types of NSs that this method has been applied to are NSs with Type I X-ray bursts, and NSs in quiescent low mass X-ray binaries (qLMXBs). Some NSs that have Type I X-ray bursts also exhibit photospheric radius expansion (PRE) bursts, and these systems have great potential  \citep{ Sztajno87,Damen90,Lewin93,Ozel06} to provide EOS constraints. Observations of PRE bursts and fitting to different spectral models has provided some estimations of the NS mass and radius \citep{Ozel09b,Guver10a,Guver10b,Suleimanov11,Poutanen14,Nattila15}. However, a variety of uncertainties in the chemical composition of the photosphere, the emission anisotropy, color correction factors, and changes in the persistent accretion flux, complicate these analyses \citep{Bhattacharyya10,Steiner10,Galloway12,Zamfir12,Worpel13,Ozel16}.

An alternative approach is to fit the emission from low-mass X-ray binaries during quiescence (qLMXBs). During quiescence, the X-rays are (often) dominated by thermal emission from the quiet NS surface, due to heating of the NS core and crust during accretion episodes \citep{Brown98}.  Nonthermal emission is often present, and typically fit by a power-law; this emission may be produced by accretion, synchrotron emission from an active pulsar wind, and/or a shock between this wind and inflowing matter \citep{Campana98,Deufel01,Cackett10,Bogdanov11,Chakrabarty14}.
The thermal emission passes through a single-component atmosphere (typically  a few cm layer of H, which would have a mass of $\sim 10^{-20}$ \Msun\ for $\sim 1$ cm \citep{Zavlin02b}) , since the elements gravitationally settle within seconds \citep{Alcock80,Hameury83}.  Current physical models of hydrogen atmospheres in low magnetic fields (appropriate for old accreting NSs) are very consistent and reliable \citep{Zavlin96,Rajagopal96,Heinke06,Haakonsen12}. 

Recent work has focused on qLMXBs in globular clusters, where the distance can be known as accurately as 6\%  \citep{Woodley12}, thus enabling stringent constraints on the radius \citep{Rutledge02a}.  Observations with {\it Chandra} and its ACIS detector (high spatial and moderate spectral resolution), or {\it XMM-Newton} with its EPIC detector (moderate spatial resolution, higher sensitivity) have allowed the identification and spectroscopy of globular cluster qLMXBs. 
Several dozen qLMXBs are now known in globular clusters, but only a few provide sufficient flux, and have sufficiently little interstellar gas absorption, to provide useful constraints \citep[e.g.][]{Heinke06,Webb07,Guillot11,Servillat12}. 
 The errors on a few of these measurements are beginning to approach 1 km, or $\sim$10\% ( see e.g. \citealt{Guillot13}), at which point they become useful for constraining nuclear physics \citep{Lattimer01}. Indeed, a new {\it Chandra} observation of the qLMXB X7 in 47 Tuc provides radius uncertainties at the 10\% level \citep{Bogdanov16}.

Thus, it has now become crucially important to identify and constrain systematic uncertainties in the qLMXB spectral fitting method.  Previous works have checked the effects of variations between hydrogen atmosphere models \citep{Heinke06,Haakonsen12}, distance errors \citep{Heinke06,Guillot11,Guillot13,Heinke14,Bogdanov16}, detector systematics \citep{Heinke06,Guillot11,Heinke14}, and modeling of the interstellar medium \citep{Heinke14,Bogdanov16}.  The largest systematic uncertainty identified so far is the atmospheric composition. If the accreted material contains no hydrogen (as expected from white dwarfs that make up 1/3 of known LMXBs in globular clusters, \citealt{Bahramian14}), then a helium (or heavier element) atmosphere will be produced.  Such helium atmospheres will have harder spectra than hydrogen atmospheres, so the inferred radii will be larger, typically by about 50\% \citep{Servillat12,Catuneanu13,Lattimer14,Heinke14}.  This uncertainty can be addressed by identification of the nature of the donor (e.g. by detecting H$\alpha$ emission, \citealt{Haggard04}, or orbital periods, \citealt{Heinke03a}).

Another serious concern is the possible presence of temperature inhomogeneities--hot spots--on the surface of the NS.  The presence of possible hot spots is a well-known concern when modeling the emission from several varieties of NSs \citep[e.g.][]{Greenstein83,Zavlin00,Pons02}.  The thermal radiation from the surface can be inhomogeneous if the polar caps of the NS are heated, either through irradiation by positrons and electrons for an active radio pulsar \citep{Harding02}, or via accretion if the magnetic field of the NS is strong enough to channel accreting matter onto the magnetic poles \citep{Gierlinski02}, or channeling of heat from the core to the poles if the internal magnetic field is of order $10^{12}$ G \citep{Greenstein83,Potekhin01,Geppert04}. The result is pulsed emission from the NS surface, which can be detected if the temperature anisotropy, spot size, geometry relative to the observer, and detector sensitivity are favorable.  Note that careful measurement of the shape of the pulse profile can constrain the ratio of mass and radius, or even both independently \citep[e.g.][]{Morsink07,Bogdanov13,Psaltis14,Miller15}; in contrast, in our case, undetected temperature inhomogeneities may bias our method.

 If the hot spots are not large or hot enough, or the emission geometry not favourable, the overall pulse amplitude may be too low to be detected. However, the undetected hot spots will affect the spectrum of the emitted light, typically hardening the spectrum compared to a star with a uniform temperature. If one were to fit the star's spectrum with a single temperature, the presence of undetected hot spots will cause the inferred temperature to be higher, and the inferred radius to be smaller, than their true values.  The fluxes from qLMXBs are generally so low that it is difficult to conduct effective pulsation searches, leaving open the possibility of hot spots.  Investigating the effect of undetected hot spots on the inferred NS radius, in the context of the qLMXBs, is the focus of this paper.

Our goal is to answer three questions.  First, what pulsed flux fraction will be produced by hot spots of relevant ranges of size and temperature difference?  Since this depends on the angle between the hot spot and NS rotational axis and between the rotational axis and the observer, the results will be probability distribution functions.   Detailed calculations for this problem have been done for blackbody emission \citep{Psaltis00,Lamb09}, with angular beaming dependence appropriate for the accretion- and nuclear-powered pulsations observed in accreting systems in outburst.  However, this calculation has not been performed specifically for hydrogen atmosphere models (which experience greater limb darkening) at temperatures relevant to quiescent NS low-mass X-ray binaries.
Second, given constraints on pulsed flux from a given quiescent NS low-mass X-ray binary, what constraints can we then impose on temperature differentials on the NS surface?  Third, how much error is incurred in calculations of the NS mass and radius by spectral fitting to a single-temperature NS, particularly for hot spots within the constraints determined above?

Although much of our calculations are general, we will apply them to the specific cases of the relatively bright ($L_X\sim10^{33}$ erg/s) quiescent NS low-mass X-ray binaries X5 and X7 in the globular cluster 47 Tuc, due to their suitability for placing constraints on the NS radius.  47 Tuc is at a distance of 4.6$\pm$0.2 kpc \citep{Woodley12,Hansen13} and experiences little Galactic reddening, $E(B-V)=0.024 \pm 0.004$ \citep{Gratton03}. X-ray emission was discovered from 47 Tuc by {\it Einstein} \citep{Hertz83}, and resolved into nine sources by {\it ROSAT} \citep{Hasinger94,Verbunt98}.  Spectral analysis of the two bright X-ray sources X5 and X7 in initial {\it Chandra} ACIS data identified them as qLMXBs with dominantly thermal X-ray emission \citep{Grindlay01a,Heinke03a}. X5 suffers varying obscuration and eclipses as a result of its edge-on 8.7-hour orbit \citep{Heinke03a}, and has a known optical counterpart \citep{Edmonds02}.  

Deeper (300 ks) {\it Chandra} ACIS observations provided large numbers of counts, enabling tight constraints on X7's radius, 14.5$^{+1.6}_{-1.4}$ km for an assumed 1.4 \Msun\ mass \citep{Heinke06}. However, these spectra suffered from significant pileup, the combination of energies from multiple X-ray photons that land in nearby pixels during one exposure \citep{Davis01}. Although a model was used to correct for this effect, this pileup model contributed unquantified systematic uncertainties to the analysis, and thus the reported constraint is no longer generally accepted \citep[e.g.][]{Steiner10}. A new, 180 ks {\it Chandra} observation of 47 Tuc in 2014-2015 was taken with {\it Chandra}'s ACIS detector in a mode minimizing pileup effects, providing a high-quality spectrum of X7 that enables tight constraints on the radius \citep{Bogdanov16}. Our simulated spectra below are designed specifically to model the effects of hot spots on this new spectrum of X7.

In addition, extremely deep (800 ks) {\it Chandra} observations of 47 Tuc have been performed with the HRC-S detector \citep{Cameron07}, which retains high (microsecond) timing resolution, though it has very poor spectral resolution.  This dataset enables a search for pulsations from X7 and X5, which we report in this work, utilizing acceleration searches \citep{Ransom01}.  Our constraints on the pulsed fractions from X7 and X5, thus, can enable us to place constraints on the effects of undetected hot spots upon their spectra.  Naturally, these constraints are probabilistic in nature, since the orientation of the NS, and of hot spots on it, affects the probability of detecting pulsations from hot spots of a given size and temperature. We also consider what constraints may be obtained from the deeper pulsation limits from {\it XMM-Newton} observations of the (non-cluster) qLMXB Cen X-4 \citet{DAngelo15}.

%====================================================================
%%%%%%%%%%%%%%%   THEORETICAL MODEL   %%%%%%%%%%%%%%  ===
%====================================================================

\section{Theoretical Model}\label{sec:pf}

\begin{figure}[t]
  \centering
  \includegraphics[scale=0.4, angle=0]{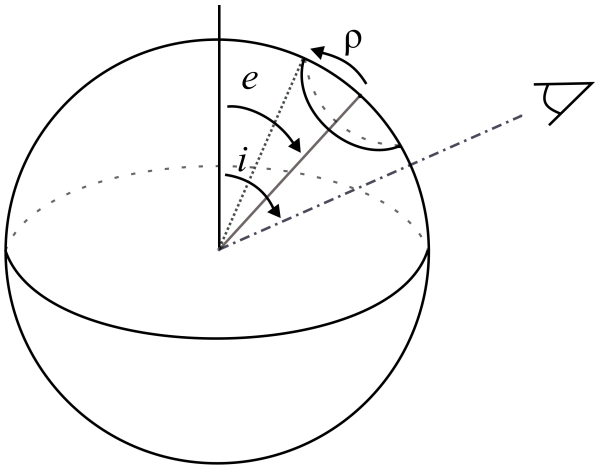}
  \caption{Schematic representation indicating the different angles. The spot's angular radius is $\rho$, and the emission angle $e$ is the  angle between the star's spin axis and the centre of the spot. The inclination angle $i$ measures the angle between the spin axis and the direction of the observer.
  \label{fig:angles}}
\end{figure}

Our model assumes a spherical neutron star of mass $M$ with radius $R$, and spin frequency $f$. The emission from most of the star is at one fixed temperature $T_{\text{NS}}$, but with one circular spot with a higher temperature $T_{spot}$. The spot's angular radius is $\rho$, and the emission angle $e$ is the  angle between the star's spin axis and the centre of the spot. The inclination angle $i$ measures the angle between the spin axis and the direction of the observer. Figure~\ref{fig:angles} shows a schematic representation of the angles used in our model. The distance to the star is $d$ and the gas column density is $N_H$. This leads to a total of 10 parameters to describe the flux from a star with a hot spot. 

For most of our calculations, we choose the spot size to match that predicted by the polar cap model,  
 \citep[][equation 18.4]{Lyne06}:
\begin{equation}\label{eq:pc}
\rho=(2\pi f R/c)^{1/2},
\end{equation}

where $c$ is the speed of light.  This formulation reduces the number of parameters in our problem by one.
This appears to be a reasonably adequate approximation for the trend of the size of X-ray emitting hot spots on radio pulsars, as suggested by phase-resolved X-ray spectral fitting of PSR J0437-4715 \citep{Bogdanov13}, the Vela pulsar \citep{Manzali07}, PSR B1055-52, and PSR B0656+14 \citep{deLuca05}.
It is not known if this is a good approximation for the spot size for qLMXBs. We will show that the dependence of pulsed fraction on spin frequency (for fixed spot size) is small.

The hydrogen atmosphere model (\citealt{McClintock04,Heinke06}; similar to that of \citealt{Zavlin96} and \citealt{Lloyd03}) assumes a thin static layer of pure hydrogen ($R_{\text{H-atm}} \ll R_{\text{NS}}$), which allows the use of a plane-parallel approximation. We assume (following e.g. \citealt{Bhattacharya91}) that the NS is weakly magnetized ($B \ll 10^{9}$ G), therefore the effects of the magnetic field on the opacity and equation of state of the atmosphere can be neglected. The opacity within the atmosphere is due to a combination of thermal free-free absorption and Thomson scattering. Light-element neutron star atmosphere models shift the peak of the emission to higher energies, relative to a blackbody model at the same effective surface temperature, due to the strong frequency dependence of free-free absorption \citep{Romani87, Zavlin96,Rajagopal96}. The opacity of the atmosphere introduces an angular dependence to the radiation which is beamed towards the normal to the surface, leading to a limb-darkening effect \citep{Zavlin96, Bogdanov07}. Limb-darkening leads to a higher pulsed fraction compared to isotropic surface emission, since the effects of light-bending and Doppler boosting are reduced \citep{Pavlov94,Bogdanov07}. 
The flux from the hydrogen atmosphere decreases slightly as the acceleration due to gravity increases, while it increases as the effective temperature increases.

The flux from the star is computed using the Schwarzschild plus Doppler approximation \citep{Miller98, Poutanen03} where the gravitational light-bending is computed using the Schwarzschild metric \citep{Pechenick83} and then Doppler effects are added as though the star were a rotating object with no gravitational field. This approximation captures the most important features of the pulsed emission for rapid rotation \citep{Cadeau07}, except for effects due to the oblateness of the star \citep{Morsink07}. The oblate shape of the star is not included in the computations done in this paper, since the oblate shape only adds small corrections to the pulsed fraction compared to factors such as the temperature differential and spot size. In addition, it has been shown \citep{Baubock15a} that the oblate shape affects the inferred radius (at the level of a few percent) for uniformly emitting blackbody stars. However, the inclusion of geometric shape effects on the inferred radius for hydrogen
atmospheres is beyond the scope of this work.  We note that these effects should be even less in hydrogen atmosphere models than in blackbody models, since the limb darkening in the hydrogen atmosphere case reduces the importance of the exact shape of the star.

In order to speed up the computations, we divide the flux calculation into three sections: $F_{\text{spot}}$, the flux from only the spot with effective temperature $T_{spot}$ (the rest of the star does not emit); $F_{\text{NS}}$, the flux from the entire star with uniform temperature $T_{\text{NS}}$; and $F_{\text{backspot}}$,  which only includes flux from the spot with effective temperature $T_{\text{NS}}$ . The total observed flux is then

\begin{equation}
F_{\text{obs}} = F_{\text{NS}} + F_{\text{spot}} - F_{\text{backspot}}, 
\end{equation}

which depends on photon energy and rotational phase. 

The computation of $F_{\text{spot}}$ is done by first choosing values for $M$, $R$, $f$, $\rho$, $i$, $e$, $T_{spot}$, $d$, and $N_H$. The Schwarzschild plus Doppler approximation is used to compute the flux at a distance $d$ from the star assuming that the parts of star outside of the spot do not emit any light. 

For the lightcurve calculation, we first calculate the X-ray absorption by the interstellar medium (using the \texttt{tbabs} model with \texttt{wilm} abundances, \citealt{Wilms00}) on the  model array, assuming $N_{H} =1.3 \times 10^{20}$ cm$^{-2}$. The $N_{H}$ is inferred from the measured $E(B-V)$ using the \citet{Predehl91} relation.   We then fold the flux model array over the \textit{Chandra} HRC effective area and a diagonal response matrix.  Finally, the flux is summed over the energy range of the detector (0 - 10 keV) for each value of rotational phase. The result is the light curve emergent from a hot spot with temperature $T_{spot}$ on the surface of a rotating neutron star, as detected by \textit{Chandra} HRC. Similarly, $F_{\text{backspot}}$ is computed in the same way, except that the effective temperature of the spot is  $T_{\text{NS}}$ instead of $T_{spot}$.  To calculate the emission  $F_{\text{NS}}$  from the entire uniformly emitting surface, we calculate the predicted flux from the \texttt{NSATMOS} model at $T_{\text{NS}}$, folded through the \texttt{tbabs} model, using the same choices of $N_H$, $M/R$ and $d$.  The pulse fraction $PF$ for the \textit{Chandra} HRC is calculated by finding the maximum and minimum values for the observed flux, and computing
\begin{equation}
PF = 
(F_{\text{obs,max}}-F_{\text{obs,min}})/ (F_{\text{obs,max}}+F_{\text{obs,min}}).
\end{equation}

To compute the expected spectrum, we follow the same procedure for calculating the observed flux ($F_{\text{obs}}$), and then integrate the flux over all phase bins at each observed energy. We then incorporate X-ray absorption by the interstellar medium, and fold the flux through the relevant \textit{Chandra} ACIS-S effective area and Response Matrix File (CALDB 4.6.3, appropriate for observations taken in 2010), ending with the phase-averaged absorbed spectrum.  \\ 

In this paper, we make reference to a fiducial star with the values of $M$ = 1.4 \Msun\ , $R = 11.5$ km, and $d = 4.6$ kpc.
We use a value for the NS effective surface temperature  $T_{\text{NS}} = 0.100$ keV (or equivalently log $T_{\text{NS}} = 6.06$), which is appropriate for the qLMXBs X5 and X7 in 47 Tuc \citep{Heinke03a,Heinke06,Bogdanov16}. 

\begin{figure}[t]
  \centering
  \includegraphics[scale=0.45]{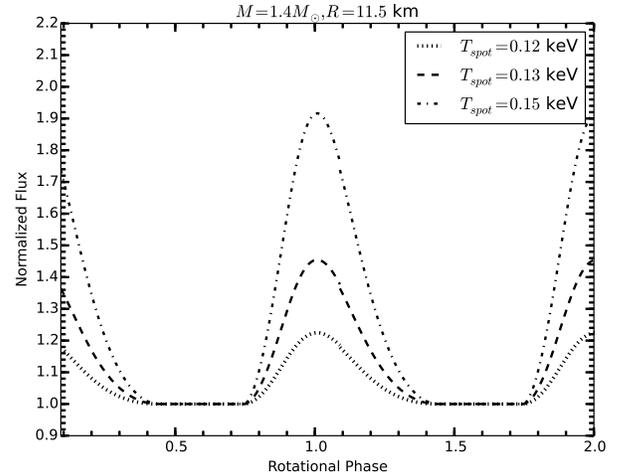}
  \caption{Pulse profiles for a 1.4 \Msun\ , 11.5 km neutron star with a hot spot at $i = 86^{\circ}$ and $e = 85^{\circ}$ at different temperature differentials. The spin frequency is 500 Hz and $\rho = 20^{\circ}$. \label{fig:profiles}}
\end{figure}

In Fig.~\ref{fig:profiles}, we show the normalised pulse profiles for the fiducial star  with different surface temperature differentials. The star spins with frequency 500 Hz, which corresponds to a spot angular radius of $20^\circ$ in Equation \ref{eq:pc}. The spot's centre is at colatitude $85^\circ$ and the observer's inclination angle is $86^\circ$. Naturally, there is a  strong dependence of the pulse fraction on the temperature of the spot. The pulse fraction increases from 16\% to 27\% when the temperature differential increases by 0.02 keV, from $T_{spot}$ = 0.13 keV to  $T_{spot}$ = 0.15 keV.

%=========================================================================
%%%%%%%%%%%%%%%   LIMITS ON PULSE FRACTION   %%%%%%%%%%%%%% ===
%=========================================================================

\section{Limits on Pulse Fraction}
\label{sec:limits}

We now address the limits on the surface temperature differentials that can be made from observational upper limits on a neutron star's pulse fraction. To investigate this, we choose different parameters $M$, $R$, $f$, $\rho$ and $T_{spot}$ describing the neutron star and its spot (with $T_{\text{NS}} = 0.1$ keV, $N_{H} =1.3 \times 10^{20}$ cm$^{-2}$ fixed for all models). 
For each choice of these parameters we then simulate the pulse profiles using the methods described in Section~\ref{sec:pf} for 300 choices of inclination, $i$, and emission, $e$, angles. We select $i$ and $e$ from distribution uniform in $\cos i$, appropriate for random orientations on the sky, and for most of our analyses, a distribution uniform in $\cos e$, random positions of the magnetic axis on the neutron star. Distributions of an angle that are uniform in the cosine of the angle tend to favour inclinations close to 90$^{\circ}$, which produce relatively large pulse fractions. We note that our assumption of a distribution of $e$, uniform in $\cos e$, may not be correct, if accreting NSs tend to shift their magnetic poles close to their rotational poles, as suggested in some theories \citep{Chen93,Chen98,Lamb09}.  Radio polarization studies do not find clear results for millisecond pulsars \citep{Manchester04}, but there is evidence from gamma-ray lightcurve fitting \citep[e.g.][]{Johnson14} and phase-resolved X-ray spectroscopy \citep[e.g.][]{Bogdanov13} that radio millisecond pulsars (descendants of LMXBs) generally have relatively large angles between their magnetic and rotational poles. To explore the effects of differing assumptions about the distribution of $e$, in our last analysis (on effects of spots on the inferred neutron star radius) we consider both a distribution uniform in $\cos e$, and one that is uniform in $e$.

\begin{figure}[t]
  \centering
  \includegraphics[scale=0.45]{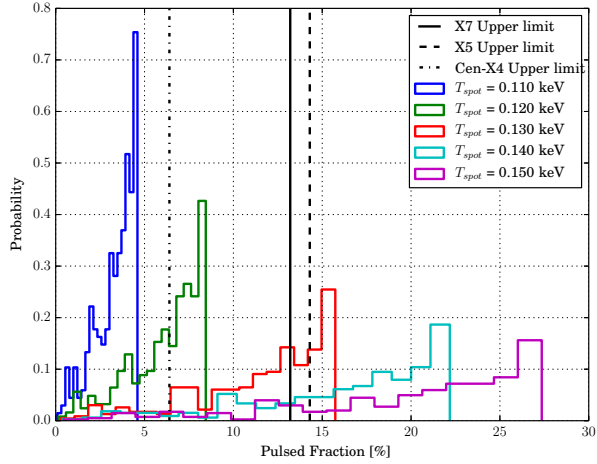}
  \caption{Histograms of simulated pulsed fractions for the fiducial NS with 300 different combinations of $i$ and $e$ for 5 different temperature differentials. The spin frequency is fixed at 500 Hz and the spot angular radius is $\rho = 20^{\circ}$.
  The neutron star surface's effective temperature is fixed at 0.10 keV with $M$ = 1.4 \Msun\ and $R=11.5$ km. 
\label{fig:vart}}
\end{figure}

We computed pulse fractions, using the same model used to generate Figure \ref{fig:profiles}, with values of $T_{spot}$ ranging from 0.105 keV to 0.160 keV, and a distribution of 300 choices of $i$ and $e$ for each spot temperature. 
In Fig.~\ref{fig:vart} we plot histograms of the pulse fractions for each  value of  the spot temperature. 

\begin{figure}[t]
  \centering
  \includegraphics[scale=0.45]{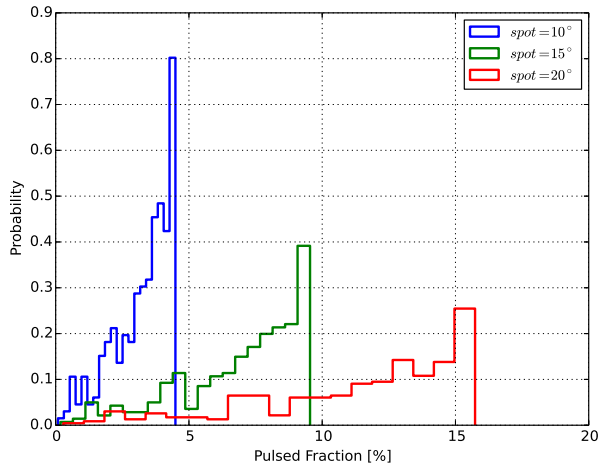}
  \caption{ Effect of angular spot radius on the histogram of pulse fractions for 300 values of $i$ and $e$. For each histogram the neutron star parameters were fixed at 
  $M$ = 1.4 \Msun\ , $R=11.5$ km, $T_{\text{NS}}=0.10$ keV,  $T_{spot}$ = 0.13 keV, $f=500$ Hz.
\label{fig:hist-rho}}
\end{figure}

The peak for each distribution corresponds to choices of $i$ and $e$ being close to $90^{\circ}$, which give the highest pulse fraction, while the tail of the  $i$ and $e$ distributions extend to $3^{\circ}$ with very small probability. As expected, as the temperature differential between the spot and the rest of the star increases, the typical pulse fraction increases, while a tail of low pulsed fraction simulations is always present. Similarly, the spot size correlates strongly with pulse fraction. In Fig. \ref{fig:hist-rho} we vary spot size, while keeping all other parameters constant. Here, Equation (\ref{eq:pc}) was not used to relate spin frequency and polar cap size, instead keeping the frequency fixed.  In Table~\ref{tab:pf} we show the 90th percentile upper \& lower limits on pulse fraction for a wide range of angular spot sizes $\rho$ and spot temperatures.

\begin{deluxetable}{llllll}
\tabletypesize{\footnotesize}
\tablecolumns{4} 
\tablewidth{0pt} 
\tablecaption{Upper and lower limits on pulse fractions for a 1.4 \Msun\ , 11.5 km neutron star at effective surface temperature 0.100 keV (Log$T= 6.06$)}
\tablehead{\colhead{$T_{spot}$}                                    &
           \colhead{$\rho$}                &
           \colhead{$f$}                             &
           \multicolumn{2}{c}{PF}                                    \\
           \colhead{[keV]}                                          &        
           \colhead{[$^{\circ}$]}                         &
           \colhead{[Hz]} &
           \colhead{$90\% <$}                                         &               
           \colhead{$90\% >$}                                                                         
               }
\startdata
    0.105 & 20 & 500 & 2.3 & 0.9  \\ 
    0.110 & 20 & 500 & 4.4 & 1.8  \\ 
    0.115 & 20 & 500 & 6.4 &  2.5 \\ 
    0.120 & 20 & 500 & 8.2 & 3.3  \\ 
    0.125 & 20 & 500 & 11.6 & 4.8  \\ 
    0.130 & 20 & 500 & 15.4 & 6.6   \\ 
    0.135 & 20 & 500 & 18.7 & 7.9   \\ 
    0.140 & 20 & 500 & 21.7 & 9.2   \\ 
    0.145 & 20 & 500 & 24.4 & 10.3   \\ 
    0.150 & 20 & 500 & 26.8 & 11.3 \\
    0.155 & 20 & 500 & 30.1 & 12.6 \\
    0.160 & 20 & 500 & 34.8 & 14.5 \\
         \cline{1-5} \\
    0.105 & 24 & 716 & 3.7 & 1.4  \\ 
    0.110 & 24 & 716 & 7.0 & 2.6  \\ 
    0.115 & 24 & 716 & 9.9 &  3.8 \\ 
    0.120 & 24 & 716 & 12.6 & 4.9  \\ 
    0.125 & 24 & 716 & 17.3 & 6.8  \\ 
    0.130 & 24 & 716 & 22.5 & 9.0   \\ 
    0.135 & 24 & 716 & 26.9 & 11.0   \\ 
    0.140 & 24 & 716 & 30.7 & 12.7   \\ 
    0.145 & 24 & 716 & 33.9 & 14.1   \\  
    0.150 & 24 & 716 & 37.3 & 15.5 \\
    0.155 & 24 & 716 & 41.1 & 17.1 \\
    0.160 & 24 & 716 & 46.3 & 19.3 \\
\enddata
    \label{tab:pf}
     \tablecomments{%
  Results for Monte Carlo simulations of 300 choices of $i$ and $e$ (drawn from distributions uniform in $\cos i$ and $\cos e$), for each choice of spot temperature and rotation rate. The spot size is determined by the polar cap model. The last two right columns represents the upper and lower 90\% bounds on the pulsed fraction.}
\end{deluxetable}

\begin{figure}[t]
  \centering
 \includegraphics[scale=0.45]{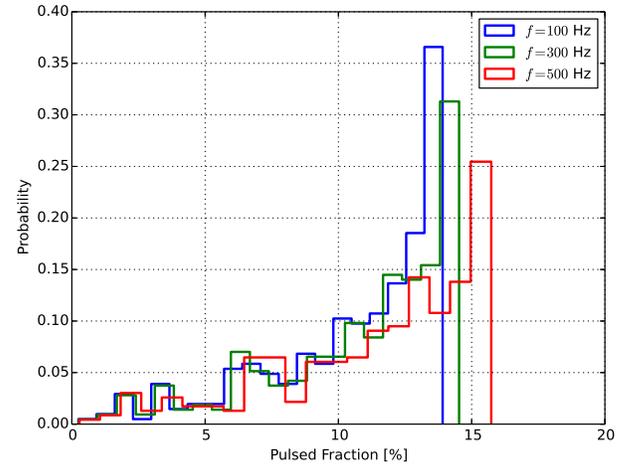}
  \caption{ Effect of spin frequency on the histogram of pulse fractions for 300 values of $i$ and $e$. For each histogram the neutron star parameters were fixed at 
  $M$ = 1.4 \Msun\ , $R=11.5$ km, $T_{\text{NS}}=0.10$ keV,  $T_{spot}$ =0.13 keV, $\rho=20^\circ$. 
\label{fig:hist-dop}}
\end{figure}

We now explore the importance of the polar cap model, Equation (\ref{eq:pc}), linking the angular spot size to the spin frequency. First, consider the effect of choosing the spin frequency independent of the spot size. As the star's spin increases, the Doppler boosting increases, which increases the intensity of the blueshifted side of the star, which will increase the pulse fraction. This effect is shown in Figure \ref{fig:hist-dop}, where it can be seen that increasing the star's spin frequency does increase the pulse fraction. However, the effect is quite small, since the pulse fraction increases only by 2\% when the frequency increases from 100 to 500 Hz. This should be contrasted with Figure \ref{fig:hist-rho} where the effect of changing the spot size but keeping the spin frequency fixed is shown. Increasing the angular spot radius by a factor of two increases the maximum pulse fraction by a factor of three. 

\begin{figure}[!ht]
  \centering
    \includegraphics[scale=0.45]{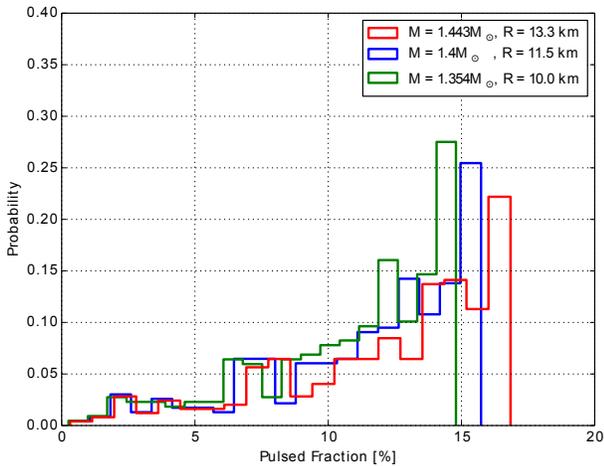}
  \caption{ Effect of $M/R$ on the histogram of pulse fractions for 300 values of $i$ and $e$. The choices of $M/R$ values are 0.16, 0.18, and 0.2 for the red, blue and green histograms respectively. For each histogram the neutron star parameters were fixed at $T_{\text{NS}}=0.10$ keV,  $T_{spot}$ = 0.13 keV, $f=500$ Hz, and $\rho=20^\circ$. Values of mass and radius are chosen so that $\log g = 14.244$. 
\label{fig:hist-mr}}
\end{figure}

\begin{figure}[t]
  \centering
 \includegraphics[scale=0.45]{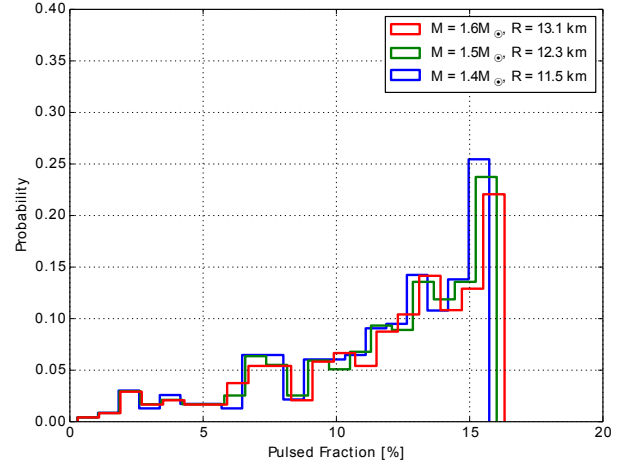}
  \caption{ Effect of surface gravity on the histogram of pulse fractions for 300 values of $i$ and $e$. The choices of $\log g$ are 14.186, 14.214 and 14.244 for the red, blue and green histograms. For each histogram the neutron star parameters were fixed at $M/R=0.18$ , $T_{\text{NS}}=0.10$ keV,  $T_{spot}$ = 0.13 keV, $f=500$ Hz, and $\rho=20^\circ$.
\label{fig:hist-g}}
\end{figure}

The choice of mass and radius affects the pulse profile through two physical effects. First, the ratio of $M/R$ controls the angles through which the light rays are bent. Larger $M/R$ gives a more compact star, which produces more gravitational bending. This leads to more of  the star being visible at any time, which produces a lower pulse fraction \citep{Pechenick83}, as can be seen in Figure \ref{fig:hist-mr}. Secondly, increasing the surface gravity (where $g=GMR^{-2}/\sqrt{1-2GM/Rc^{2}}$) alters the emission pattern, decreasing the limb darkening, which decreases the pulse fraction.  The effect of the surface gravity is shown in Figure~\ref{fig:hist-g} where different values of $M$ and $R$ are chosen so that the ratio $M/R$ is kept constant. The largest star has the lowest surface gravity and the largest pulse fraction. Both effects are small, with changes in $M/R$ causing changes in the pulse fraction of a similar order as the changes due to spin frequency. The effects due to surface gravity changes are even smaller. All of these effects act to increase the pulse fraction if the radius of the star is increased while keeping the mass constant, as shown
by \citet{Bogdanov07}. 

In this work, we simulate the effects of one spot. Adding a second spot would typically reduce the measured pulse fraction. This depends on the compactness of the star, on the angles $e$ and $i$, and on whether the spots are antipodal.  For angles $e$ and $i$ near 90 degrees, one spot will always be visible (for typical neutron star compactness values, such as our 11.5 km, 1.4 \Msun\ standard star), which will reduce the pulsed fraction. However, if the angles $e$ and $i$ are both far from 90 degrees, then the far spot will not be strongly visible and the pulsed fraction will not change dramatically.  Thus, the effect on the histograms of pulsed fractions will be to shift the peak to smaller values, but the tail at low values (which is made up of realizations with small values of $e$ and/or $i$) will be much less affected. 
The 90th percentile lower limits on the pulse fraction are set by the tail at low values, so the pulsed fraction lower limits will generally not be strongly affected by adding a second spot (assuming it is antipodal to the first spot).

%=======================================================================
%%%%%%%%%%%%%%%   APPLICATIONS on qLMXBs   %%%%%%%%%%%%%% ===
%=======================================================================

\subsection{Application to qLMXBs in 47 Tuc, M28 and Cen X-4}
\label{s:pf-app}

\begin{deluxetable}{llllll}
\tabletypesize{\footnotesize}
\tablecolumns{4} 
\tablewidth{0pt} 
\tablecaption{\textit{Chandra} HRC archival data of globular cluster qLMXBs.}
\tablehead{\colhead{Cluster/source}                                    &
           \colhead{ObsID}                &
           \colhead{Date}                             &
           \colhead{Exposure (ks)}                                    \\                                                                  
               }
\startdata
47 Tucanae  &	5542	& 2005 Dec 19   &	50.16	\\
X5  \& X7   &	5543	& 2005 Dec 20	&	51.39 \\
	&	5544	& 2005 Dec 21 	&	50.14	\\
	&	5545	& 2005 Dec 23	&	51.87	\\
	&	5546	& 2005 Dec 27	&	50.15	\\
	&	6230	& 2005 Dec 28 	&	49.40	\\
	&	6231	& 2005 Dec 29	&	47.15	\\
	&	6232	& 2005 Dec 31	&	44.36	\\
	&	6233	& 2006 Jan 2 	&	97.93	\\
	&	6235	& 2006 Jan 4 	&	50.13	\\
	&	6236	& 2006 Jan 5 	&	51.92	\\
	&	6237	& 2005 Dec 24	&	50.17	\\
	&	6238	& 2005 Dec 25	&	48.40	\\
	&	6239	& 2006 Jan 6	&	50.16	\\
	&	6240	& 2006 Jan 8	&	49.29	\\
	\hline
		&		&			&			\\
M28    	&	2797	& 2002 Nov 8   &	49.37	\\
Source 26	&	6769	& 2006 May 27 &  	41.07	\\
\enddata
  \label{tab:obs}
\end{deluxetable}

Among globular cluster qLMXBs with thermal spectra, only three (X5 and X7 in 47 Tuc, and source 26 in M28) have substantial observations with a telescope and instrument with the timing and spatial resolution (Chandra's HRC-S camera in timing mode) to conduct significant searches for pulsations at spin periods of milliseconds.\footnote{Note that \citet{Papitto13} searched for pulsations from the accreting millisecond X-ray pulsar IGR J18245-2452 during an intermediate-luminosity ($1.4\times10^{33}$ erg/s) outburst, using a 53-ks HRC-S observation of M28 and a known ephemeris for the pulsar, and placed an upper limit of 17\% on the pulse amplitude.} These targets have not previously been searched for pulsations.

We extracted lightcurves from X7 and X5 from 800 ksec of {\it Chandra} HRC-S data, obtained during December  2005 to January 2006, described in \citep{Cameron07}. To search for pulsations from qLXMBs we make use of \textit{Chandra} HRC-S observations, which offer  a time resolution of $\sim$16 $\mu$s in the special SI mode. We extracted source events for the 47 Tuc qLMXBs X7 and X5 from mutiple HRC-S exposures acquired in 2005 and 2006 (see \citealt{Cameron07}) and the M28 qLMXB (named Source 26 by \citealt{Becker03}) from two exposures  obtained in 2002 and 2006 \citep{Rutledge04, Bogdanov11}. Table~\ref{tab:obs} summarizes the archival observations that were used in this analysis.  For each source the events were extracted from circular regions of radius 2.5$''$ centered on the positions obtained from {\tt wavdetect}. The recorded arrival times were then translated to the solar system barycenter using the {\tt axbary} tool in CIAO assuming the DE405 solar system ephemeris. The HRC provides no reliable spectral information so all collected events were used for the analysis below. 

The pulsation searches were conducted using the PRESTO pulsar search software package. Given that NS qLMXBs are by definition in compact binaries, the detection of X-ray pulsations from these objects in blind  periodicity searches is complicated by the binary motion of the NS, which smears out the pulsed signal over numerous Fourier bins and thus diminishes its detectability.  Therefore, it is necessary to employ Fourier-domain periodicity search techniques that compensate for the binary motion when searching for spin-induced flux variations. For this analysis, we use two complementary methods: acceleration searches and ``sideband'' (or phase-modulation) modulation searches. For the acceleration search technique, the algorithm attempts to recover the loss of power caused by the large period derivative induced by the  rapid orbital motion \citep{Ransom02}. This method is most effective when the exposure time of the observation is a small fraction of the orbital period. In contrast, the sideband technique is most effective when the observation is much longer than the orbital period, provided that the observation is contiguous \citep{Ransom03}. This approach identifies sidebands produced in the power spectrum centered around the intrinsic spin period and stacks them in order to recover some sensitivity to the pulsed signal.   

The orbital periods of LMXBs are typically  of order hours, or for the case of ultracompact systems, $\lesssim$1 hour. Due to the relatively low count rates of the three qLMXB sources, searching for pulsations over short segments of the binary orbit ($\lesssim$30 minutes)  is not feasible so acceleration searches are insensitive to pulsations from these targets.  As the \textit{Chandra} exposures are longer than the orbital cycle of X5 and likely for X7 and M28 source 26 as well,  the phase-modulation method is the most effective for this purpose.

The maximum frequency that we search up to sets our number of trials, and thus sets how strong an upper limit we can set. The pulse fraction limit increases as we go to higher frequencies because it is necessary to bin the event data for the acceleration and sideband searches. This causes frequency dependent attenuation of the signal - resulting in decreased sensitivity at high frequencies  \citep[e.g.][]{Middleditch76,Leahy83}. The pulse fraction upper limits were obtained in PRESTO, which considers the maximum power found in the power spectrum as described in \citet{Vaughan94}. We find no evidence for coherent X-ray pulsations in any of the individual observations of the three qLMXBs. The most restrictive  upper limits on the X-ray pulsed fraction were obtained from the longest exposures. We find that for spin periods as low as 2 ms (500 Hz), the 90\% upper limit on any pulsed signal 14\%, 13\%, and 37\%, for X5, X7 and M28 source 26.

Performing searches up to the fastest known neutron star spin period \citep{Hessels06}, 1.4 ms (716 Hz), the limits are 16\%, 15\%, and 37\%. Since the pulsed fraction upper limit for Source 26 in M28 is so high, it does not lead to useful constraints, so we do not consider it further in our analysis.

Our pulsed fraction upper limit on X7 (as an example) places limits on the temperature differentials that the NS may have. For a spin frequency of 500 Hz, the 90\% upper limit of 13\% can be compared with the pulse fraction probabilities for different spot temperatures shown in Table \ref{tab:pf}. For example, for a spot temperature of 0.125 keV, 90\% of computed models have a pulse fraction smaller than 11.6\%. In fact, all computed models at this spot temperature (0.125 keV) have pulse fractions below the 13\% upper limit for X7, so this temperature differential is consistent with the observations. This means that X7 could have an undetected hot spot. However, increasing the spot temperature to 0.130 keV, the histogram plotted in Figure \ref{fig:vart} shows that only 58\% of our simulations give a pulsed fraction below the 90\% upper limit on X7's pulsed fraction. For higher spot temperatures, it becomes more improbable to have an undetected hot spot; that is, a pulse fraction below the 90\% upper limit on the pulse fraction for X7. We find that a spot temperature of 0.155 keV, or a temperature differential of 0.055 keV, to be the maximum temperature differential allowable for X7. This calculation assumes that X7 is spinning at 500 Hz. 
 For higher spin frequencies, which give a larger spot radius (as we linked frequency to spot radius), we get higher pulsed fractions when other inputs are identical. Therefore, the maximum temperature differential allowable slightly decreases to 0.050 keV (spot temperature of 0.150 keV) above which, over 90\% of the simulations are above the 90\% pulse fraction upper limit of X7. The 90\% upper limit on X5's pulsed fraction is only 1\% larger than that for X7, which will increase the maximum temperature differential allowable for X5 by a few percent more than allowed for X7 ($\sim$0.005 keV larger).
Our computations all assume that the neutron star has $M$ = 1.4 \Msun\ and $R=11.5$ km, however, our results show that the dependence on mass and radius is weak. 
 
Even more stringent constraints are possible from the accreting neutron star in Cen X-4, which was observed at a similar luminosity as X7, but at a distance of only 1.2 kpc \citep{Chevalier89}, with a more sensitive X-ray telescope, {\it XMM-Newton}. \citet{DAngelo15} used a deep (80 ks) \textit{XMM-Newton} observation \citep{Chakrabarty14}, in which the PN camera was operated in timing mode (with 30 $\mu$s time resolution), to search for pulsations. D'Angelo et al. utilized a semicoherent search strategy, in which short segments of data are searched coherently, and then combined incoherently \citep{Messenger11}.  This analysis assumed a circular orbit, with orbital period and semimajor axis as measured by \citet{Chevalier89}, but left orbital phase free. D'Angelo et al. calculated a fractional-amplitude upper limit of 6.4\% from Cen X-4 in quiescence. This is significantly lower than the pulsed fraction limits in X5 and X7, so it provides a tighter constraint on the temperature differential, as can be seen in Fig.~\ref{fig:vart}.  If we assume the neutron star in Cen X-4 to have the same physical properties as X7, then its maximum spot temperature must be smaller.  Our simulations show that even with this small upper limit Cen X-4 can have small temperature differentials (up to 0.01 keV) with all simulations being below the pulsed fraction upper limit.  The maximum spot temperature the neutron star in Cen X-4 may have is 0.130 keV, at which 90\% of the simulations have pulsed fractions that are above the 90\% upper limit. Similarly, for higher frequencies, at 716 Hz, the maximum allowable spot temperature decreases to only 0.125 keV.
Next, we address how these possible hot spots could affect spectroscopic inferences of neutron star radii, and what limits we can place on these effects from our constraints on the pulsed fraction and temperature differential.

%==================================================================
%%%%%%%%%%%%%%%   SPECTRAL EFFECTS   %%%%%%%%%%%%%% ===
%==================================================================

\section{Effect of a Hot Spot on the Spectrum}
\label{s:spectrum}

The existence of a hot spot causes a change in the observed spectrum. To illustrate the effect we choose an extreme case, corresponding to our fiducial star rotating at 500 Hz spin frequency, plus a spot with $T_{spot} = 0.15$ keV,
angular radius $\rho = 20^{\circ}$, and emission and inclination angles $e = 85^{\circ}$ and $i = 86^{\circ}$. The pulse profile for this case is shown in Figure \ref{fig:profiles} and has a pulsed fraction of 31\%. 
The method described in Section~\ref{sec:pf} is used to compute the flux from the spot and the rest of the star. We compute the spectra for each rotational phase of the neutron star over the energy range (0.2 - 10.0 keV), then we integrate the spectra over all rotational phases to produce the simulated phase-averaged spectrum. We convolve the flux from the spot and star with our interstellar medium model, then fold them over the proper response matrix and effective area of the \Chandra\, ACIS-S detector. We fix the exposure time in our simulation at 200 ks (chosen to represent the 2014--2015 \Chandra\ / ACIS observation of 47 Tuc), then use a Poisson distribution to select the number of counts per energy bin.

\begin{figure}[t]
  \centering
  \includegraphics[scale=0.45, angle=0]{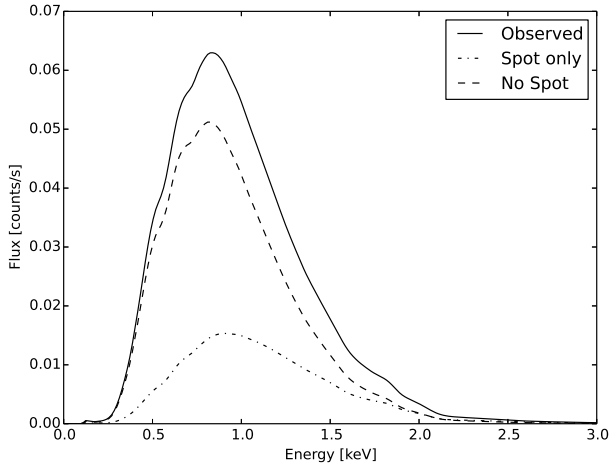}
  \caption{The effect of the existence of hot spots on the observed spectrum. The neutron star has a surface temperature of 0.10 keV, and the hot spot is at 0.15 keV. The peak of the spectrum slightly shifts to a higher energy by 0.02 keV. The hotter the spot is, the more distorted the spectrum will be. 
  \label{fig:spot}}
\end{figure}

 Figure~\ref{fig:spot} shows an example spectrum. The dashed curve shows the flux $F_{\text{NS}}$ integrated over phase, which corresponds to the flux from all parts of a  star at $T_{\text{NS}}=0.1$ keV.  The dotted curve shows the flux from the hot spot $F_{\text{spot}}$, at temperature 0.15 keV. The solid curve shows the observed flux $F_{\text{obs}} = F_{\text{NS}} + F_{\text{spot}} - F_{\text{backspot}}$. The peak of the observed spectrum is shifted by $\sim$ 0.02 keV, and the flux increases by over 20\%. The shift of the peak photon energy is smaller than the energy resolution of \Chandra\ / ACIS at lower energies (of order 0.1 keV).  
 
We now test how the spectroscopically inferred radius changes if the star has a hot spot, but the spectral fitting assumes that the star's emission is homogeneous.  We simulated spectra for the fiducial star with $M$ = 1.4 \Msun\ , $R=11.5$ km, $T_{\text{NS}}=0.1$ keV, $d=4.6$ kpc, $f=500$ Hz, and $N_{H}$ = 1.3 $\times 10^{20}$ cm$^{-2}$ with a hot spot on the surface. 
We chose a variety of temperature differentials and spot sizes (assuming the polar cap model), as shown in Table~\ref{tab:fit2}. For each model, we use the heasoft tool \texttt{FLX2XSP} to convert the flux array to a PHA spectrum, which we load into \texttt{XSPEC} to fit. We let $R$ and $T_{\text{NS}}$ be free in the spectral fit, while we fix the mass at $M$ = 1.4 \Msun\ and the distance $d = 4.6$ kpc. We allow $N_{H}$  to be free, but with a minimum value of $1.3 \times 10^{20}$ cm$^{-2}$.  The resulting \texttt{XSPEC} fitted values and uncertainties for the radius and temperature are shown along with the reduced chi-squared in Table~\ref{tab:fit2}. 

The first row in Table~\ref{tab:fit2} shows the uncertainty inherent in the method, by first simulating a light curve for a star with no hot spot; the best-fit radius is quite close (0.1 km) to the input value, and the radius uncertainty (0.7-0.8 km) is consistent with that from fitting to real data on X7 (Bogdanov et al. 2016). 
Next, we see that there is a systematic trend in the inferred radii of the neutron stars introduced by an undetected hot spot.  XSPEC interprets the shifted spectrum as an increase in the temperature of the whole star. However, the observed flux will not be as large as one would expect for the higher temperature, so this is interpreted as indicating a smaller star. The general result is that the star's radius is under-estimated when an undetected hot spot is present. This effect can be seen in many of the best-fit solutions shown in Table~\ref{tab:fit2}. The principal factor in introducing bias is the spot temperature; a 15\% bias in the average fitted temperature is induced by spot temperatures of 0.14 to 0.15 keV, for spot sizes between 9$^{\circ}$ and 23$^{\circ}$.  For this reason, we focus on the spot temperature  as the crucial variable to explore below.  
%%%

\begin{figure*}[ht]
    \centering
        \includegraphics[scale=0.6]{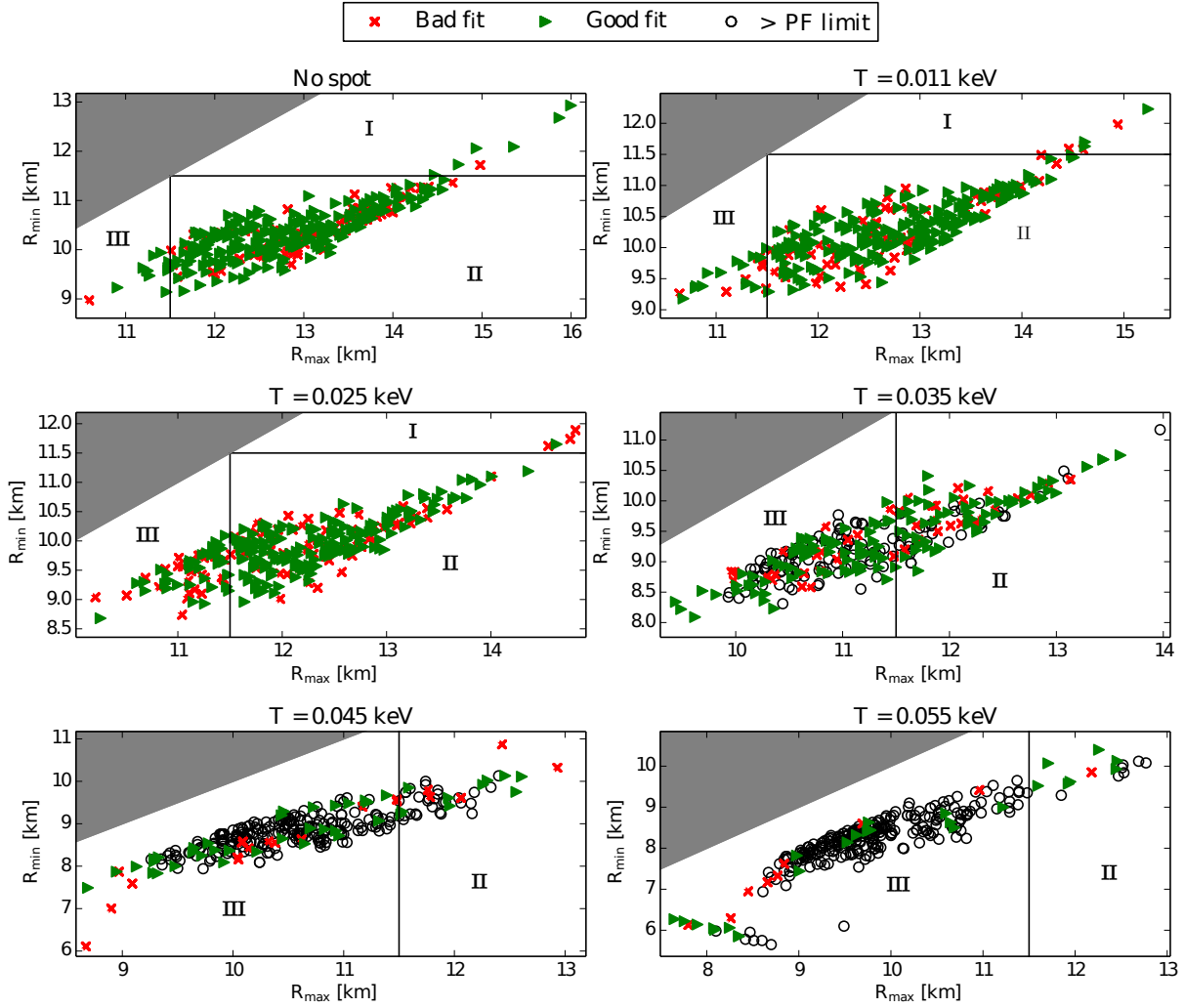}
    \caption{Calculated upper and lower radius limits (90\% confidence) from fitting 300 spectral simulations with different choices of the temperature differential, assuming a 1.4 \Msun\ NS, with the angles $e$ and $i$ chosen from distributions uniform in $\cos i$ and $\cos e$. The shaded area is prohibited, and the solid lines represent the ``true'' (input to simulation) value of the neutron star radius, $R_{\text{NS}} = 11.5$ km. Points in the lower right quadrant of each graph indicate fits where the ``true'' (input) radius falls between the inferred upper and lower radius limits, while points in the lower left quadrant show a radius upper limit below the ``true'' value.
The results shown here are directly applicable to the neutron star X7 in 47 Tuc, which has a 90\% upper limit of 13\% on the pulsed fraction. 
    }\label{fig:fit_x7}
\end{figure*}

Table~\ref{tab:fit2} only shows results for one particular choice of emission and inclination angles. For a more general picture, for each value of $T_{spot}$ we simulated 300 spectra with emission and inclination angles drawn from distributions uniform in $\cos i$ and $\cos e$ for the fiducial star, assuming a 500 Hz spin, a 1.4 \Msun\ mass, and a radius of 11.5 km. Each simulation was fit in \texttt{XSPEC} using the same method used for Table~\ref{tab:fit2}.  The resulting 90\% confidence limits on the radius from each simulation are indicated by coloured dots in Figures~\ref{fig:fit_x7}. 
 Each graph shows the results for a particular spot temperature and has four regions, separated by black lines indicating the input value of the neutron star's radius $R$ used in the simulation (the ``true'' radius). The region with $R_{\text{max}} \ge R$ and $R_{\text{min}} \le R$ (lower right-hand quadrant) corresponds to fits that are consistent with the correct radius. The points in the region with  $R_{\text{min}} > R$ (upper right-hand quadrant) are fits that overestimate the neutron star's radius, while the points in the region with $R_{\text{max}} < R$ (lower left-hand quadrant) underestimate the radius. The fourth region, shaded grey, is forbidden since it corresponds to $R_{\text{min}} > R_{\text{max}}$. 

To determine whether the spectral distortion due to a hot spot would be detectable, and thus whether NSs with hot spots might be identified by their poor fits to single-temperature models, we retained fit quality information for each fit. We define each fit with a reduced chi-squared value greater than 1.1 (which indicates a null hypothesis probability less than 0.044, given the 51 degrees of freedom) to be a ``bad'' fit, and mark it as a red cross. Unfortunately, the fraction of ``bad'' fits does not increase substantially with increasing hot spot temperature (Fig.~\ref{fig:fit_x7}, and Table~\ref{tab:fit3}), indicating that fit quality cannot effectively identify spectra with hot spots.

Each simulation also has an associated pulsed fraction. If the pulsed fraction is larger than the measured upper-limit for X7 (for an assumed spin of 500 Hz), we marked it as a black hollow circle. Good fits that do not violate the pulsed-fraction limit are marked as a green triangle.
For spot temperatures up to 0.125 keV we find that over 75\% of the simulations give inferred radii that are consistent with the true value of $R_{\text{NS}}$.
For higher temperature differentials ($T_{spot}$$>$ 0.13 keV) a large fraction of the inferred radii are biased downward from the "true" value by larger than 10\% of the true radius of the neutron star, while the majority ($>$58\%) of the simulations are below the X7 pulse fraction upper limit. This pulse fraction limit, and the inferred bias, changes if the spin frequency (and consequently the spot size) changes. For the higher spin frequency of 716 Hz we find that inferred radii can be biased up to 15\% smaller than the true radius of the neutron star for $T_{spot}$=0.13 keV. In Table ~\ref{tab:fit3}, we summarize the percentage of inferred radii consistent with the "true" value, the percentage of good fits, and the average bias in the inferred radius for different choices of spot temperature. We examined the behaviour of $(R_{\text{max}} + R_{\text{min}}) / 2$ vs. $R_{\text{fit}}$, finding a well-behaved linear relationship between the two quantities.  In this paper we calculate the bias as the difference between the median of the inferred $R_{\text{fit, no spot}}$ radii with no hot spots and the median of the inferred radii $R_{\text{fit, spot}}$ with a hot spot, divided by the latter. (This definition allows this bias to be directly applicable to observed radius estimates). The $R_{\text{fit, no spot}}$ values are results of fitting 300 simulated  spectra from a poisson distribution, which would give a distribution of $R_{\text{fit}}$ peaked at the \textit{true} value of $R=11.5$ km.

\begin{figure*}[ht]
    \centering
        \includegraphics[scale=0.45]{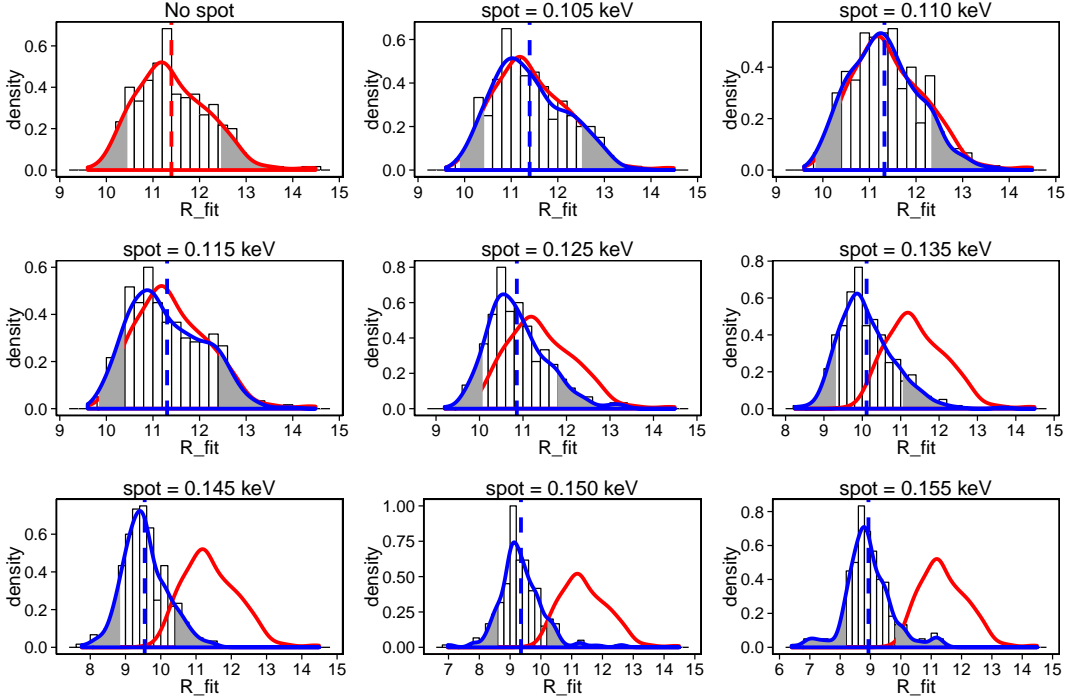}
    \caption{Distribution of ($R_{\text{fit}}$) from fitting 300 spectral simulations for different choices of the temperature differential, assuming a 1.4 \Msun\ NS, with the angles $e$ and $i$ chosen from distributions uniform in $\cos i$ and $\cos e$. The red curve is the probability density curve for the simulations without a hot spot (essentially the systematic errors inherent in the method), while the blue curve indicates the probability density of the inferred $R_{\text{fit}}$ at each hot spot temperature. The dashed line is the mean of ($R_{\text{fit}}$). The shaded grey areas exclude the upper and lower 10\% of each probability density curve. The theoretical model is for a 11.5 km neutron star spinning at 500 Hz. These histograms show the bias in radii measurements.
    }\label{fig:rhisto}
\end{figure*}

In Fig.~\ref{fig:rhisto} we present histograms of the inferred $R_{\text{fit}}$ at different spot temperatures, and compare it to the distribution of inferred $R_{\text{fit}}$ with no spot. This shows the bias in the mean between the histogram with no spot and the histogram of inferred radii with a hot spot. For a spot temperature as high as 0.125 keV, the bias in $R_{\text{fit}}$ is still at or below 5\%. 
At 0.130 keV, the majority of simulations do not violate the pulse fraction limit, the bias in the mean is 10\%, and over half the fits are consistent with the input radius. 
For the maximum spot temperature we allow in our simulations, the bias in the mean of $R_{\text{fit}}$ can reach up to 40\%, however < 10\% of the simulations at this spot temperature are below the upper limits for either X7 or Cen X-4.

\begin{figure}[t]
    \centering
        \includegraphics[scale=0.45]{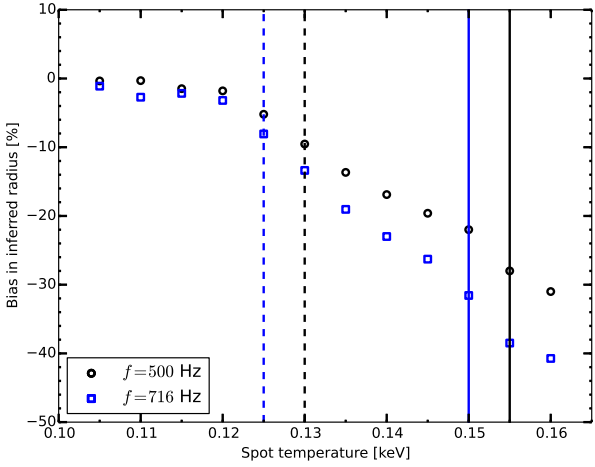}
    \caption{Bias in the spectroscopically inferred $R_{max}$ (90\% confidence) as a function of the spot temperature relative to a NS at surface temperature of 0.100 keV. The black colour is associated spinning frequency of 500 Hz and the blue colour is associated the 716 Hz. The solid and dashed lines are the maximum allowable spot temperatures that would not give rise to detectable pulsations based on the pulse fraction limits for X7 and Cen-X4 respectively.
    }\label{fig:bias}
\end{figure}

To identify a reasonable limiting case, we choose the $T_{spot}$ where less than 10\% of the simulations provide pulse fractions below the upper limit on each neutron star's pulse fraction; thus, 0.155 keV for X7, and 0.130 keV for Cen X-4. This allows a maximum downward bias in their spectroscopically inferred radii of up to 28\% for X7, and 10\% for Cen X-4. For example, if we assume the neutron star in Cen X-4 to be a 1.4 \Msun\ star spinning at 500 Hz with a  spectroscopically inferred radius of exactly 11.5 km, an undetected hot spot could allow a true radius as high as 12.65 km. For X7, in the case of maximal undetected hot spots, the measured radius of 11.1$^{+0.8}_{-0.7}$ km (for an assumed 1.4 \Msun\ neutron star mass, Bogdanov et al. 2016) could allow a true radius up to 15.2 km in the extreme case.   In Fig. ~\ref{fig:bias} we summarize the bias in $R_{\text{fit}}$ versus spot temperature at 500 Hz and 716 Hz frequencies. At both frequencies, the bias is below 10\%  for relatively small spot temperatures (up to 0.125 keV). 
However, at higher spot temperatures (> 0.130 keV) there is a clear divergence between the magnitude of the biases at 500 Hz and 716 Hz, becoming larger with spot temperature. 
Increasing the frequency from 500 to 716 Hz changes the bias from 32\% to 41\% at the highest spot temperature (0.160 keV), but since the 716 Hz frequency also has a larger pulsed fraction for the same spot temperature, the maximum spot temperature is reduced in the 716 Hz case, and the actual maximum bias in the 500 and 716 Hz cases is similar. Finally, we ran the simulations with choices of a uniform distribution of $e$ and $\cos i$. This produces lower pulsed fractions when compared to simulations using a uniform distribution of $\cos e$ at the same spot temperature (see numbers in parentheses in Table.\ref{tab:fit3}) .  In turn,  this increases the maximum allowable spot temperature that would not give rise to detectable pulsations. For X7, the maximum spot temperature for an assumed uniform distribution of $e$ is larger than 0.160 keV (the limit of our model), while the maximum spot temperature for Cen X-4 would be 0.155 keV, both at the spin frequency of 500 Hz.  A hot spot temperature larger than 0.160 keV would give a spectroscopically inferred radius less than 50\% of the true radius, which essentially means the bias is not usefully bounded.

\begin{deluxetable}{cccr|ccc}[!th]
  \tablecolumns{7}
  \tablecaption{%
    Best-fit values for $R$
    \label{tab:fit2} 
  }%   
  \tablehead{%
    \colhead{$T_{spot}$}             &
    \colhead{$\rho$}       &
    \colhead{$f$}            &
    \colhead{PF} &
    \colhead{$R_{\text{fit}}$}   &
    \colhead{Log$T_{\text{eff,fit}}$}   &
    \colhead{$\chi^{2}_{\nu}$} \\
    \colhead{[keV]}      &
    \colhead{$[^{\circ}]$}   &
    \colhead{[Hz]}    &
    \colhead{[\%]}   &
    \colhead{[km]}  &
    \colhead{} &
    \colhead{}
  }
  \startdata 
      ...  & ... & ... &   ... & 11.4$^{+0.8}_{-0.7}$ & 6.05$^{+0.02}_{-0.02}$ & $0.99$ \\ 
     \cline{1-7} \\
     0.13 & 9  & 100 &  3.3& 11.4$^{+1.3}_{-0.8}$ & 6.04$^{+0.02}_{-0.02}$ & $1.09$  \\ 
     0.15 & 9  & 100 &  9.0 & 9.7$^{+1.1}_{-0.6}$ & 6.09$^{+0.02}_{-0.02}$ & $1.09$  \\
     \cline{1-7} \\
    0.11 & 20 & 500 & 5.5 & 11.8$^{+1.3}_{-0.8}$ & 6.03$^{+0.03}_{-0.02}$ & $1.10$   \\ 
    0.12 & 20 & 500 & 9.8 & 11.6$^{+0.9}_{-0.8}$ & 6.04$^{+0.02}_{-0.02}$ & $0.82$ \\ 
    0.13 & 20 & 500 &18.0 & 10.9$^{+0.8}_{-0.8}$ & 6.07$^{+0.02}_{-0.02}$ & $1.15$ \\ 
    0.14 & 20 & 500 & 25.2 & 9.8$^{+0.8}_{-0.6}$ & 6.10$^{+0.02}_{-0.02}$ & $1.04$ \\ 
    0.15 & 20 & 500 & 30.8 & 9.3$^{+0.8}_{-0.8}$ & 6.11$^{+0.02}_{-0.02}$ & $1.01$  \\ 
     \cline{1-7} \\
    0.11 & 23 & 667 & 7.7 & 11.6$^{+0.8}_{-0.8}$ & 6.04$^{+0.02}_{-0.02}$ & $1.06$  \\ 
    0.12 & 23 & 667 &13.9 & 10.9$^{+0.8}_{-0.6}$ & 6.06$^{+0.02}_{-0.02}$ & $0.70$ \\      
    0.13 & 23 & 667 &24.5 & 10.9$^{+1.1}_{-0.8}$ & 6.07$^{+0.02}_{-0.03}$ & $0.98$ \\ 
    0.14 & 23 & 667 & 33.2 & 9.1$^{+0.9}_{-0.6}$ & 6.13$^{+0.02}_{-0.02}$ & $1.18$ \\ 
    0.15 & 23 & 667 &39.5 & 8.8$^{+1.0}_{-0.4}$ & 6.14$^{+0.03}_{-0.02}$ & $1.40$  \\     
    \enddata
     \tablecomments{%
Best-fit values of $R$ and $T_{eff}$ for given choices of $T_{spot}$, spot size $\rho$, spin frequency, and constant angles $i$ = 80$^{\circ}$ and $e$ = 89$^{\circ}$.   The spectra are generated assuming $M$ = 1.4 \Msun\ , $R = 11.5$ km, surface temperature $T_{\text{NS}}$=0.10 keV, $\log T_{\text{NS}}=6.06$. Errors are 90\% confidence.  Spectral fits assume $M$ = 1.4 \Msun\ and $d = 4.6$ kpc.  The pulse fractions produced by each simulation are provided for reference.
    }
\end{deluxetable}

\begin{deluxetable}{lllllll}[!th]
  \tablecolumns{8}
  \tablecaption{%
   \label{tab:fit3} 
  }%   
  \tablehead{%
    \colhead{$T_{spot}$} 	&
    \colhead{$f$}       			&
    \colhead{< X7 limit}       		&
    \colhead{< Cen-X4 limit}       	&
    \colhead{Consistent}       	&
     \colhead{Good fits}        	&
    \colhead{Bias}  			\\
    \colhead{[keV]}      		&
    \colhead{[Hz]}   			&
    \colhead{[\%]}    			&
    \colhead{[\%]}    			&
    \colhead{[\%]}    			&
    \colhead{[\%]}    			&
    \colhead{[\%]}
  }
  \startdata 
    0.105	&	500		&	100	 (100)	&	100	 (100)	&	93 	&	79	&	$-\, 0.3$	\\ 
    0.110 	& 	500 	& 	100 (100)	&  	100 (100)	&	92 	& 	75	&	$-\, 0.4$	\\ 
    0.115 	& 	500 	& 	100 (100)	& 	87	(89)	&	90  	& 	78 	&	$-\, 1$	\\ 	
    0.120 	& 	500 	& 	100 (100)	&	45	(56)	&	84  	& 	73	&	$-\, 2$	\\ 
    0.125 	& 	500 	& 	100 (100)	& 	22	(34)	&	76  	& 	69	&	$-\, 5	$	\\ 	
    0.130 	& 	500 	& 	58  	(64)	& 	10	(26)	&	55  	& 	73	&	$-\, 10$	\\    
    0.135 	& 	500 	& 	33  	(47)	& 	8	(20)	&	38  	& 	72	&	$-\, 14$	\\ 
    0.140 	& 	500 	& 	24  	(38)	& 	7	(16)	&	19	& 	71	&	$-\, 17$	\\ 
    0.145 	& 	500 	& 	21  	(33)	& 	6	(14)	&	15 	& 	69	&	$-\, 20$	\\  
    0.150 	& 	500 	& 	17 	(31)	&   	5	(12)	&	9 	& 	76	&	$-\, 22$	\\
    0.155 	& 	500 	& 	12 	(28)	&   	3	(10)	&	5	& 	71	&	$-\, 28$	\\
    0.160 	& 	500 	& 	8	(23)	&  	2	(8)		&	0.6	& 	57	&	$-\, 32$	\\	
      	  	&  		&	  	&  			&			&					\\ 
    0.105 	& 	716 	& 	100 	&	100	&	89 	& 	77	&	$-\, 1$	\\ 
    0.110 	& 	716 	& 	100 	& 	73	&	87 	& 	79	&	$-\, 2$	\\ 
    0.115 	& 	716 	& 	100 	& 	32	&	86 	&	78	&	$-\, 3$	\\ 	
    0.120 	& 	716 	& 	100 	& 	21	&	79 	& 	79 	&	$-\, 3$	\\ 
    0.125 	& 	716 	&  	 63 	& 	9	&	62 	&	79	&	$-\, 8$	\\ 	
    0.130 	& 	716 	&  	 32 	& 	7	&	33 	&	77	&	$-\, 13$	\\    
    0.135 	& 	716 	&  	 22	& 	6	&	16 	&	74	&	$-\, 19$	\\ 
    0.140 	& 	716 	&  	 17 	& 	4	&	10 	&	72 	&	$-\, 23$	\\ 
    0.145 	& 	716 	&  	 13 	& 	3	&	5  	&	67	&	$-\, 26$	\\ 
    0.150 	& 	716 	& 	10 	&   	3	&	4 	& 	65	&	$-\, 31$	\\
    0.155 	& 	716 	& 	8 	&   	2	&	2	& 	56	&	$-\, 38$	\\
    0.160 	& 	716 	& 	7	&  	2	&	0.6	& 	57	&	$-\, 41$	\\
          	& 	 	& 	 	& 		&		&		&			\\
    \enddata
     \tablecomments{%
For different spot temperatures, the bias (right column) in radius determinations, and the percentages of simulations that lie under the upper limits on the pulsed fraction for X7 and Cen X-4, that give spectral fits consistent with the ``true'' radius, and that give ``good'' fits ($\chi^2_{\nu}$<1.1).
 Each line gives results from fitting 300 simulated spectra using $R = 11.5$ km and surface temperature $T_{\text{NS}}$=0.10 keV, for different choices of spot temperatures and spin frequency. Spectral fits assume $M$ = 1.4 \Msun\ and $d = 4.6$ kpc.  
The percentage of ``good fits'' are the percentage of the simulations below the upper limits. Numbers in brackets are for simulations performed with a uniform distribution of $e$ (rather than uniform in $\cos e$).
    }
\end{deluxetable}

\subsection{Limits of our analysis}

Our analysis necessarily is limited in scope. Here, we enumerate some complexities that we have not addressed in this work. 
The temperature distribution of the hot spots may be more complex than we have assumed; especially for large spots, this might cause significant changes \citep[see, e.g.][]{Baubock15b}.  We have sampled only a few values of the spin period, mass, and radius. We have assumed hydrogen atmospheres; helium atmospheres, while generally similar in spectra and angular dependences, have some subtle differences \citep[see ][ figures 5 and 9]{Zavlin96}. These issues are unlikely to significantly alter our results.

A larger issue is that we assume that the neutron star has only one spot. A second spot would reduce the average pulsed fraction, though it would probably not reduce the lower limit on the pulsed fraction substantially (see section 3). The second spot would generally increase the visible amount of the star at a higher temperature, so it would increase the bias in the radius. Some NSs have strong evidence for poles that are not offset by 180 degrees \citep[e.g.][]{Bogdanov13}, and/or with different sizes and temperatures \citep{Gotthelf10}, adding additional possible complexity.

Another major issue is that the distribution of $e$ may not be uniform in either $\cos e$ or in $e$; if hot spots are more concentrated towards the poles than we assume (as suggested by \citealt{Lamb09}), then the pulsed fractions will tend to be lower than we assume. 

A final issue, relating to the applicability of our results to other systems, is that our simulations were designed with surface temperature and extinction ($N_H$) designed to match specific qLMXBs in 47 Tuc. Increased $N_H$ would tend to obscure the softer emission from the full surface more than the hot spot, thus increasing the expected pulse fraction and the expected bias in spectral fitting. 

%==============================================================
%%%%%%%%%%%%%%%   CONCLUSIONS   %%%%%%%%%%%%%% ===
%==============================================================

\section{Conclusion}

We studied the effects of hot spots on the X-ray lightcurves, spectra, and spectroscopically inferred masses and radii, for neutron stars with hydrogen atmospheres. Hydrogen atmospheres, due to limb darkening, display higher pulsed fractions than blackbody emission, so this analysis is necessary in order to constrain the systematic effects of radius measurements on quiescent neutron stars.
We find that the existence of an unmodeled hot spot tends to shift the peak to higher energies, which affects the spectroscopically inferred equatorial radii of neutron stars. 

We first computed the 90\% upper limits on the pulsed fractions from 800 ks \Chandra\ HRC-S observation for the two sources X5 and X7 in the globular cluster 47-Tuc to be 14\% and 13\% respectively, searching spin frequencies < 500 Hz. For higher spin frequencies (up to 716 Hz) the limits are 16\% and 15\% respectively. 
We simulated pulse profiles for ranges of inclination and hot spot emission angles $i$ and $e$ , obtaining the central 90\% range of pulse fraction obtained for different choices of temperature differentials (between the hot spot and the rest of the NS) and NS spin frequencies. This allows us to constrain the maximum temperature differential for any hot spots on X5 and X7. In the case of X7, if we assume it is a 1.4 \Msun\ neutron star spinning at 500 Hz, our results indicate that the maximum allowable temperature differential  is 0.055 keV, where > 90\% of our simulations are above the 90\% upper limit of pulsed fraction. The neutron star in Cen-X4 has a significantly lower upper limit on the pulse fraction of 6.4\%, which puts a tighter constraint on the maximum allowable temperature differential of 0.025 keV. Since the upper limit of Source 26 in M28 is high (37\%), it does not provide strong constraints.

Finally, we study the effects on the inferred radius of hot spots for these temperature differential limits. The spectroscopically inferred radii of stars with spots tend to be at smaller values than the ``true'' radius. The 90\% confidence range of the inferred radii are generally still consistent with the true value of our fiducial star (11.5 km) for small temperature differentials (0.03 keV). 

For the hottest possible hot spots that would not give rise to detectable pulsations in X7, we find that a bias in the best-fit inferred radius of up to 28\% smaller than the true radius may be induced by hot spots below our upper limit. For Cen X-4 (where the pulse fraction constraint is much tighter, $<$6.4\%), downward radius biases are constrained to < 10\%.  If the hot spot emission angle $e$ is distributed uniformly in $e$ (rather than in $\cos e$, as appropriate if the hot spot may be anywhere on the neutron star surface), then the constraints are significantly looser, and effectively unbounded for the X7 case. 
 Our analysis constrains a key systematic uncertainty in the most promising radius measurement method.  We do not know whether quiescent neutron stars in X-ray binaries without radio pulsar activity have hot spots.  However, the possibility strongly motivates further pulsation searches in quiescent neutron stars in X-ray binaries, particularly  those that are targets for spectroscopic radius determination.

%===================================================================
%%%%%%%%%%%%%%%   ACKNOWLEDGMENTS   %%%%%%%%%%%%%% ===
%===================================================================

\acknowledgments
The authors are grateful to M.~C. Miller and S. Guillot for discussions and comments on the draft, and to S. Ransom for discussions and for providing the PRESTO pulsation search code. 
COH has been supported by an NSERC Discovery Grant, an Ingenuity New Faculty Award, and an Alexander von Humboldt Fellowship, and thanks the Max Planck Institute for Radio Astronomy in Bonn for their hospitality. SMM has been supported by an NSERC Discovery Grant.

\bibliographystyle{apj}
\bibliography{Bibliography}

\begin{thebibliography}{}
\expandafter\ifx\csname natexlab\endcsname\relax\def\natexlab#1{#1}\fi

\bibitem[{{Alcock} \& {Illarionov}(1980)}]{Alcock80}
{Alcock}, C., \& {Illarionov}, A. 1980, \apj, 235, 534

\bibitem[{{Antoniadis} {et~al.}(2013){Antoniadis}, {Freire}, {Wex}, {Tauris},
  {Lynch}, {van Kerkwijk}, {Kramer}, {Bassa}, {Dhillon}, {Driebe}, {Hessels},
  {Kaspi}, {Kondratiev}, {Langer}, {Marsh}, {McLaughlin}, {Pennucci}, {Ransom},
  {Stairs}, {van Leeuwen}, {Verbiest}, \& {Whelan}}]{Antoniadis13}
{Antoniadis}, J., {Freire}, P.~C.~C., {Wex}, N., {et~al.} 2013, Science, 340,
  448

\bibitem[{{Bahramian} {et~al.}(2014){Bahramian}, {Heinke}, {Sivakoff},
  {Altamirano}, {Wijnands}, {Homan}, {Linares}, {Pooley}, {Degenaar}, \&
  {Gladstone}}]{Bahramian14}
{Bahramian}, A., {Heinke}, C.~O., {Sivakoff}, G.~R., {et~al.} 2014, \apj, 780,
  127

\bibitem[{{Baub{\"o}ck} {et~al.}(2015{\natexlab{a}}){Baub{\"o}ck}, {{\"O}zel},
  {Psaltis}, \& {Morsink}}]{Baubock15a}
{Baub{\"o}ck}, M., {{\"O}zel}, F., {Psaltis}, D., \& {Morsink}, S.~M.
  2015{\natexlab{a}}, \apj, 799, 22

\bibitem[{{Baub{\"o}ck} {et~al.}(2015{\natexlab{b}}){Baub{\"o}ck}, {Psaltis},
  \& {{\"O}zel}}]{Baubock15b}
{Baub{\"o}ck}, M., {Psaltis}, D., \& {{\"O}zel}, F. 2015{\natexlab{b}}, \apj,
  811, 144

\bibitem[{{Becker} {et~al.}(2003){Becker}, {Swartz}, {Pavlov}, {Elsner},
  {Grindlay}, {Mignani}, {Tennant}, {Backer}, {Pulone}, {Testa}, \&
  {Weisskopf}}]{Becker03}
{Becker}, W., {Swartz}, D.~A., {Pavlov}, G.~G., {et~al.} 2003, \apj, 594, 798

\bibitem[{{Bhattacharya} \& {van den Heuvel}(1991)}]{Bhattacharya91}
{Bhattacharya}, D., \& {van den Heuvel}, E.~P.~J. 1991, \physrep, 203, 1

\bibitem[{{Bhattacharyya}(2010)}]{Bhattacharyya10}
{Bhattacharyya}, S. 2010, Advances in Space Research, 45, 949

\bibitem[{{Bogdanov}(2013)}]{Bogdanov13}
{Bogdanov}, S. 2013, \apj, 762, 96

\bibitem[{{Bogdanov} {et~al.}(2011){Bogdanov}, {Archibald}, {Hessels}, {Kaspi},
  {Lorimer}, {McLaughlin}, {Ransom}, \& {Stairs}}]{Bogdanov11}
{Bogdanov}, S., {Archibald}, A.~M., {Hessels}, J.~W.~T., {et~al.} 2011, \apj,
  742, 97

\bibitem[{{Bogdanov} {et~al.}(2016){Bogdanov}, {Heinke}, {{\"O}zel}, \&
  {G{\"u}ver}}]{Bogdanov16}
{Bogdanov}, S., {Heinke}, C.~O., {{\"O}zel}, F., \& {G{\"u}ver}, T. 2016, ArXiv
  e-prints, arXiv:1603.01630

\bibitem[{{Bogdanov} {et~al.}(2007){Bogdanov}, {Rybicki}, \&
  {Grindlay}}]{Bogdanov07}
{Bogdanov}, S., {Rybicki}, G.~B., \& {Grindlay}, J.~E. 2007, \apj, 670, 668

\bibitem[{{Brown} {et~al.}(1998){Brown}, {Bildsten}, \& {Rutledge}}]{Brown98}
{Brown}, E.~F., {Bildsten}, L., \& {Rutledge}, R.~E. 1998, \apjl, 504, L95

\bibitem[{{Cackett} {et~al.}(2010){Cackett}, {Brown}, {Miller}, \&
  {Wijnands}}]{Cackett10}
{Cackett}, E.~M., {Brown}, E.~F., {Miller}, J.~M., \& {Wijnands}, R. 2010,
  \apj, 720, 1325

\bibitem[{{Cadeau} {et~al.}(2007){Cadeau}, {Morsink}, {Leahy}, \&
  {Campbell}}]{Cadeau07}
{Cadeau}, C., {Morsink}, S.~M., {Leahy}, D., \& {Campbell}, S.~S. 2007, \apj,
  654, 458

\bibitem[{{Cameron} {et~al.}(2007){Cameron}, {Rutledge}, {Camilo}, {Bildsten},
  {Ransom}, \& {Kulkarni}}]{Cameron07}
{Cameron}, P.~B., {Rutledge}, R.~E., {Camilo}, F., {et~al.} 2007, \apj, 660,
  587

\bibitem[{{Campana} {et~al.}(1998){Campana}, {Stella}, {Mereghetti}, {Colpi},
  {Tavani}, {Ricci}, {Dal Fiume}, \& {Belloni}}]{Campana98}
{Campana}, S., {Stella}, L., {Mereghetti}, S., {et~al.} 1998, \apjl, 499, L65

\bibitem[{{Catuneanu} {et~al.}(2013){Catuneanu}, {Heinke}, {Sivakoff}, {Ho}, \&
  {Servillat}}]{Catuneanu13}
{Catuneanu}, A., {Heinke}, C.~O., {Sivakoff}, G.~R., {Ho}, W.~C.~G., \&
  {Servillat}, M. 2013, \apj, 764, 145

\bibitem[{{Chakrabarty} {et~al.}(2014){Chakrabarty}, {Tomsick}, {Grefenstette},
  {Psaltis}, {Bachetti}, {Barret}, {Boggs}, {Christensen}, {Craig},
  {F{\"u}rst}, {Hailey}, {Harrison}, {Kaspi}, {Miller}, {Nowak}, {Rana},
  {Stern}, {Wik}, {Wilms}, \& {Zhang}}]{Chakrabarty14}
{Chakrabarty}, D., {Tomsick}, J.~A., {Grefenstette}, B.~W., {et~al.} 2014,
  \apj, 797, 92

\bibitem[{{Chen} \& {Ruderman}(1993)}]{Chen93}
{Chen}, K., \& {Ruderman}, M. 1993, \apj, 408, 179

\bibitem[{{Chen} {et~al.}(1998){Chen}, {Ruderman}, \& {Zhu}}]{Chen98}
{Chen}, K., {Ruderman}, M., \& {Zhu}, T. 1998, \apj, 493, 397

\bibitem[{{Chevalier} \& {Ilovaisky}(1989)}]{Chevalier89}
{Chevalier}, C., \& {Ilovaisky}, S.~A. 1989, in ESA Special Publication, Vol.
  296, Two Topics in X-Ray Astronomy, Volume 1: X Ray Binaries. Volume 2: AGN
  and the X Ray Background, ed. J.~{Hunt} \& B.~{Battrick}, 345--347

\bibitem[{{Damen} {et~al.}(1990){Damen}, {Magnier}, {Lewin}, {Tan}, {Penninx},
  \& {van Paradijs}}]{Damen90}
{Damen}, E., {Magnier}, E., {Lewin}, W.~H.~G., {et~al.} 1990, \aap, 237, 103

\bibitem[{{D'Angelo} {et~al.}(2015){D'Angelo}, {Fridriksson}, {Messenger}, \&
  {Patruno}}]{DAngelo15}
{D'Angelo}, C.~R., {Fridriksson}, J.~K., {Messenger}, C., \& {Patruno}, A.
  2015, \mnras, 449, 2803

\bibitem[{{Davis}(2001)}]{Davis01}
{Davis}, J.~E. 2001, \apj, 562, 575

\bibitem[{{De Luca} {et~al.}(2005){De Luca}, {Caraveo}, {Mereghetti},
  {Negroni}, \& {Bignami}}]{deLuca05}
{De Luca}, A., {Caraveo}, P.~A., {Mereghetti}, S., {Negroni}, M., \& {Bignami},
  G.~F. 2005, \apj, 623, 1051

\bibitem[{{Demorest} {et~al.}(2010){Demorest}, {Pennucci}, {Ransom}, {Roberts},
  \& {Hessels}}]{Demorest10}
{Demorest}, P.~B., {Pennucci}, T., {Ransom}, S.~M., {Roberts}, M.~S.~E., \&
  {Hessels}, J.~W.~T. 2010, \nat, 467, 1081

\bibitem[{{Deufel} {et~al.}(2001){Deufel}, {Dullemond}, \& {Spruit}}]{Deufel01}
{Deufel}, B., {Dullemond}, C.~P., \& {Spruit}, H.~C. 2001, \aap, 377, 955

\bibitem[{{Edmonds} {et~al.}(2002){Edmonds}, {Heinke}, {Grindlay}, \&
  {Gilliland}}]{Edmonds02}
{Edmonds}, P.~D., {Heinke}, C.~O., {Grindlay}, J.~E., \& {Gilliland}, R.~L.
  2002, \apjl, 564, L17

\bibitem[{{Freire} {et~al.}(2011){Freire}, {Bassa}, {Wex}, {Stairs},
  {Champion}, {Ransom}, {Lazarus}, {Kaspi}, {Hessels}, {Kramer}, {Cordes},
  {Verbiest}, {Podsiadlowski}, {Nice}, {Deneva}, {Lorimer}, {Stappers},
  {McLaughlin}, \& {Camilo}}]{Freire11}
{Freire}, P.~C.~C., {Bassa}, C.~G., {Wex}, N., {et~al.} 2011, \mnras, 412, 2763

\bibitem[{{Galloway} \& {Lampe}(2012)}]{Galloway12}
{Galloway}, D.~K., \& {Lampe}, N. 2012, \apj, 747, 75

\bibitem[{{Geppert} {et~al.}(2004){Geppert}, {K{\"u}ker}, \&
  {Page}}]{Geppert04}
{Geppert}, U., {K{\"u}ker}, M., \& {Page}, D. 2004, \aap, 426, 267

\bibitem[{{Gierli{\'n}ski} {et~al.}(2002){Gierli{\'n}ski}, {Done}, \&
  {Barret}}]{Gierlinski02}
{Gierli{\'n}ski}, M., {Done}, C., \& {Barret}, D. 2002, \mnras, 331, 141

\bibitem[{{Gotthelf} {et~al.}(2010){Gotthelf}, {Perna}, \&
  {Halpern}}]{Gotthelf10}
{Gotthelf}, E.~V., {Perna}, R., \& {Halpern}, J.~P. 2010, \apj, 724, 1316

\bibitem[{{Gratton} {et~al.}(2003){Gratton}, {Bragaglia}, {Carretta}, \& {et
  al.}}]{Gratton03}
{Gratton}, R.~G., {Bragaglia}, A., {Carretta}, E., \& {et al.} 2003, \aap, 408,
  529

\bibitem[{{Greenstein} {et~al.}(1983){Greenstein}, {Dolez}, \&
  {Vauclair}}]{Greenstein83}
{Greenstein}, J.~L., {Dolez}, N., \& {Vauclair}, G. 1983, \aap, 127, 25

\bibitem[{{Grindlay} {et~al.}(2001){Grindlay}, {Heinke}, {Edmonds}, \&
  {Murray}}]{Grindlay01a}
{Grindlay}, J.~E., {Heinke}, C., {Edmonds}, P.~D., \& {Murray}, S.~S. 2001,
  Science, 292, 2290

\bibitem[{{Guillot} {et~al.}(2011){Guillot}, {Rutledge}, {Brown}, {Pavlov}, \&
  {Zavlin}}]{Guillot11}
{Guillot}, S., {Rutledge}, R.~E., {Brown}, E.~F., {Pavlov}, G.~G., \& {Zavlin},
  V.~E. 2011, \apj, 738, 129

\bibitem[{{Guillot} {et~al.}(2013){Guillot}, {Servillat}, {Webb}, \&
  {Rutledge}}]{Guillot13}
{Guillot}, S., {Servillat}, M., {Webb}, N.~A., \& {Rutledge}, R.~E. 2013, \apj,
  772, 7

\bibitem[{{G{\"u}ver} {et~al.}(2010{\natexlab{a}}){G{\"u}ver}, {{\"O}zel},
  {Cabrera-Lavers}, \& {Wroblewski}}]{Guver10b}
{G{\"u}ver}, T., {{\"O}zel}, F., {Cabrera-Lavers}, A., \& {Wroblewski}, P.
  2010{\natexlab{a}}, \apj, 712, 964

\bibitem[{{G{\"u}ver} {et~al.}(2010{\natexlab{b}}){G{\"u}ver}, {Wroblewski},
  {Camarota}, \& {{\"O}zel}}]{Guver10a}
{G{\"u}ver}, T., {Wroblewski}, P., {Camarota}, L., \& {{\"O}zel}, F.
  2010{\natexlab{b}}, \apj, 719, 1807

\bibitem[{{Haakonsen} {et~al.}(2012){Haakonsen}, {Turner}, {Tacik}, \&
  {Rutledge}}]{Haakonsen12}
{Haakonsen}, C.~B., {Turner}, M.~L., {Tacik}, N.~A., \& {Rutledge}, R.~E. 2012,
  \apj, 749, 52

\bibitem[{{Haensel} {et~al.}(2016){Haensel}, {Bejger}, {Fortin}, \&
  {Zdunik}}]{Haensel16}
{Haensel}, P., {Bejger}, M., {Fortin}, M., \& {Zdunik}, L. 2016, European
  Physical Journal A, 52, 59

\bibitem[{{Haggard} {et~al.}(2004){Haggard}, {Cool}, {Anderson}, {Edmonds},
  {Callanan}, {Heinke}, {Grindlay}, \& {Bailyn}}]{Haggard04}
{Haggard}, D., {Cool}, A.~M., {Anderson}, J., {et~al.} 2004, \apj, 613, 512

\bibitem[{{Hameury} {et~al.}(1983){Hameury}, {Heyvaerts}, \&
  {Bonazzola}}]{Hameury83}
{Hameury}, J.~M., {Heyvaerts}, J., \& {Bonazzola}, S. 1983, \aap, 121, 259

\bibitem[{{Hansen} {et~al.}(2013){Hansen}, {Kalirai}, {Anderson}, {Dotter},
  {Richer}, {Rich}, {Shara}, {Fahlman}, {Hurley}, {King}, {Reitzel}, \&
  {Stetson}}]{Hansen13}
{Hansen}, B.~M.~S., {Kalirai}, J.~S., {Anderson}, J., {et~al.} 2013, \nat, 500,
  51

\bibitem[{{Harding} {et~al.}(2002){Harding}, {Strickman}, {Gwinn}, {Dodson},
  {Moffet}, \& {McCulloch}}]{Harding02}
{Harding}, A.~K., {Strickman}, M.~S., {Gwinn}, C., {et~al.} 2002, \apj, 576,
  376

\bibitem[{{Hasinger} {et~al.}(1994){Hasinger}, {Johnston}, \&
  {Verbunt}}]{Hasinger94}
{Hasinger}, G., {Johnston}, H.~M., \& {Verbunt}, F. 1994, \aap, 288, 466

\bibitem[{{Hebeler} {et~al.}(2013){Hebeler}, {Lattimer}, {Pethick}, \&
  {Schwenk}}]{Hebeler13}
{Hebeler}, K., {Lattimer}, J.~M., {Pethick}, C.~J., \& {Schwenk}, A. 2013,
  \apj, 773, 11

\bibitem[{{Heinke} {et~al.}(2003){Heinke}, {Grindlay}, {Lloyd}, \&
  {Edmonds}}]{Heinke03a}
{Heinke}, C.~O., {Grindlay}, J.~E., {Lloyd}, D.~A., \& {Edmonds}, P.~D. 2003,
  \apj, 588, 452

\bibitem[{{Heinke} {et~al.}(2006){Heinke}, {Rybicki}, {Narayan}, \&
  {Grindlay}}]{Heinke06}
{Heinke}, C.~O., {Rybicki}, G.~B., {Narayan}, R., \& {Grindlay}, J.~E. 2006,
  \apj, 644, 1090

\bibitem[{{Heinke} {et~al.}(2014){Heinke}, {Cohn}, {Lugger}, {Webb}, {Ho},
  {Anderson}, {Campana}, {Bogdanov}, {Haggard}, {Cool}, \&
  {Grindlay}}]{Heinke14}
{Heinke}, C.~O., {Cohn}, H.~N., {Lugger}, P.~M., {et~al.} 2014, \mnras, 444,
  443

\bibitem[{{Hertz} \& {Grindlay}(1983)}]{Hertz83}
{Hertz}, P., \& {Grindlay}, J.~E. 1983, \apj, 275, 105

\bibitem[{{Hessels} {et~al.}(2006){Hessels}, {Ransom}, {Stairs}, {Freire},
  {Kaspi}, \& {Camilo}}]{Hessels06}
{Hessels}, J.~W.~T., {Ransom}, S.~M., {Stairs}, I.~H., {et~al.} 2006, Science,
  311, 1901

\bibitem[{{Johnson} {et~al.}(2014){Johnson}, {Venter}, {Harding}, {Guillemot},
  {Smith}, {Kramer}, {{\c C}elik}, {den Hartog}, {Ferrara}, {Hou}, {Lande}, \&
  {Ray}}]{Johnson14}
{Johnson}, T.~J., {Venter}, C., {Harding}, A.~K., {et~al.} 2014, \apjs, 213, 6

\bibitem[{{Lamb} {et~al.}(2009){Lamb}, {Boutloukos}, {Van Wassenhove},
  {Chamberlain}, {Lo}, {Clare}, {Yu}, \& {Miller}}]{Lamb09}
{Lamb}, F.~K., {Boutloukos}, S., {Van Wassenhove}, S., {et~al.} 2009, \apj,
  706, 417

\bibitem[{{Lattimer} \& {Prakash}(2001)}]{Lattimer01}
{Lattimer}, J.~M., \& {Prakash}, M. 2001, \apj, 550, 426

\bibitem[{{Lattimer} \& {Prakash}(2007)}]{Lattimer07}
---. 2007, \physrep, 442, 109

\bibitem[{{Lattimer} \& {Prakash}(2016)}]{Lattimer16}
---. 2016, \physrep, 621, 127

\bibitem[{{Lattimer} \& {Steiner}(2014)}]{Lattimer14}
{Lattimer}, J.~M., \& {Steiner}, A.~W. 2014, \apj, 784, 123

\bibitem[{{Leahy} {et~al.}(1983){Leahy}, {Darbro}, {Elsner}, {Weisskopf},
  {Kahn}, {Sutherland}, \& {Grindlay}}]{Leahy83}
{Leahy}, D.~A., {Darbro}, W., {Elsner}, R.~F., {et~al.} 1983, \apj, 266, 160

\bibitem[{{Lewin} {et~al.}(1993){Lewin}, {van Paradijs}, \& {Taam}}]{Lewin93}
{Lewin}, W.~H.~G., {van Paradijs}, J., \& {Taam}, R.~E. 1993, \ssr, 62, 223

\bibitem[{{Lloyd}(2003)}]{Lloyd03}
{Lloyd}, D.~A. 2003, ArXiv e-prints, astro-ph/0303561

\bibitem[{Lyne {et~al.}(2006)Lyne, Graham-Smith, \& Graham-Smith}]{Lyne06}
Lyne, A., Graham-Smith, F., \& Graham-Smith, F. 2006, Pulsar Astronomy,
  Cambridge Astrophysics (Cambridge University Press)

\bibitem[{{Manchester} \& {Han}(2004)}]{Manchester04}
{Manchester}, R.~N., \& {Han}, J.~L. 2004, \apj, 609, 354

\bibitem[{{Manzali} {et~al.}(2007){Manzali}, {De Luca}, \&
  {Caraveo}}]{Manzali07}
{Manzali}, A., {De Luca}, A., \& {Caraveo}, P.~A. 2007, \apj, 669, 570

\bibitem[{{McClintock} {et~al.}(2004){McClintock}, {Narayan}, \&
  {Rybicki}}]{McClintock04}
{McClintock}, J.~E., {Narayan}, R., \& {Rybicki}, G.~B. 2004, \apj, 615, 402

\bibitem[{{Messenger}(2011)}]{Messenger11}
{Messenger}, C. 2011, \prd, 84, 083003

\bibitem[{{Middleditch}(1976)}]{Middleditch76}
{Middleditch}, J. 1976, PhD thesis, California Univ., Berkeley.

\bibitem[{{Miller} \& {Lamb}(1998)}]{Miller98}
{Miller}, M.~C., \& {Lamb}, F.~K. 1998, \apjl, 499, L37

\bibitem[{Miller \& Miller(2015)}]{Miller15}
Miller, M.~C., \& Miller, J.~M. 2015, Phys. Rep., 548, 1

\bibitem[{{Morsink} {et~al.}(2007){Morsink}, {Leahy}, {Cadeau}, \&
  {Braga}}]{Morsink07}
{Morsink}, S.~M., {Leahy}, D.~A., {Cadeau}, C., \& {Braga}, J. 2007, \apj, 663,
  1244

\bibitem[{{N{\"a}ttil{\"a}} {et~al.}(2015){N{\"a}ttil{\"a}}, {Steiner},
  {Kajava}, {Suleimanov}, \& {Poutanen}}]{Nattila15}
{N{\"a}ttil{\"a}}, J., {Steiner}, A.~W., {Kajava}, J.~J.~E., {Suleimanov},
  V.~F., \& {Poutanen}, J. 2015, ArXiv e-prints, arXiv:1509.06561

\bibitem[{{{\"O}zel}(2006)}]{Ozel06}
{{\"O}zel}, F. 2006, \nat, 441, 1115

\bibitem[{{\"O}zel(2013)}]{Ozel13}
{\"O}zel, F. 2013, Reports on Progress in Physics, 76, 016901

\bibitem[{{{\"O}zel} {et~al.}(2009){{\"O}zel}, {G{\"u}ver}, \&
  {Psaltis}}]{Ozel09b}
{{\"O}zel}, F., {G{\"u}ver}, T., \& {Psaltis}, D. 2009, \apj, 693, 1775

\bibitem[{{{\"O}zel} {et~al.}(2016){{\"O}zel}, {Psaltis}, {G{\"u}ver}, {Baym},
  {Heinke}, \& {Guillot}}]{Ozel16}
{{\"O}zel}, F., {Psaltis}, D., {G{\"u}ver}, T., {et~al.} 2016, \apj, 820, 28

\bibitem[{{Papitto} {et~al.}(2013){Papitto}, {Ferrigno}, {Bozzo}, {Rea},
  {Pavan}, {Burderi}, {Burgay}, {Campana}, {di Salvo}, {Falanga},
  {Filipovi{\'c}}, {Freire}, {Hessels}, {Possenti}, {Ransom}, {Riggio},
  {Romano}, {Sarkissian}, {Stairs}, {Stella}, {Torres}, {Wieringa}, \&
  {Wong}}]{Papitto13}
{Papitto}, A., {Ferrigno}, C., {Bozzo}, E., {et~al.} 2013, \nat, 501, 517

\bibitem[{{Pavlov} {et~al.}(1994){Pavlov}, {Shibanov}, {Ventura}, \&
  {Zavlin}}]{Pavlov94}
{Pavlov}, G.~G., {Shibanov}, Y.~A., {Ventura}, J., \& {Zavlin}, V.~E. 1994,
  \aap, 289, 837

\bibitem[{{Pechenick} {et~al.}(1983){Pechenick}, {Ftaclas}, \&
  {Cohen}}]{Pechenick83}
{Pechenick}, K.~R., {Ftaclas}, C., \& {Cohen}, J.~M. 1983, \apj, 274, 846

\bibitem[{{Pons} {et~al.}(2002){Pons}, {Walter}, {Lattimer}, {Prakash},
  {Neuh{\"a}user}, \& {An}}]{Pons02}
{Pons}, J.~A., {Walter}, F.~M., {Lattimer}, J.~M., {et~al.} 2002, \apj, 564,
  981

\bibitem[{{Potekhin} \& {Yakovlev}(2001)}]{Potekhin01}
{Potekhin}, A.~Y., \& {Yakovlev}, D.~G. 2001, \aap, 374, 213

\bibitem[{{Poutanen} \& {Gierli{\'n}ski}(2003)}]{Poutanen03}
{Poutanen}, J., \& {Gierli{\'n}ski}, M. 2003, \mnras, 343, 1301

\bibitem[{{Poutanen} {et~al.}(2014){Poutanen}, {N{\"a}ttil{\"a}}, {Kajava},
  {Latvala}, {Galloway}, {Kuulkers}, \& {Suleimanov}}]{Poutanen14}
{Poutanen}, J., {N{\"a}ttil{\"a}}, J., {Kajava}, J.~J.~E., {et~al.} 2014,
  \mnras, 442, 3777

\bibitem[{{Predehl} {et~al.}(1991){Predehl}, {Hasinger}, \&
  {Verbunt}}]{Predehl91}
{Predehl}, P., {Hasinger}, G., \& {Verbunt}, F. 1991, \aap, 246, L21

\bibitem[{{Psaltis} {et~al.}(2014){Psaltis}, {{\"O}zel}, \&
  {Chakrabarty}}]{Psaltis14}
{Psaltis}, D., {{\"O}zel}, F., \& {Chakrabarty}, D. 2014, \apj, 787, 136

\bibitem[{{Psaltis} {et~al.}(2000){Psaltis}, {{\"O}zel}, \&
  {DeDeo}}]{Psaltis00}
{Psaltis}, D., {{\"O}zel}, F., \& {DeDeo}, S. 2000, \apj, 544, 390

\bibitem[{{Rajagopal} \& {Romani}(1996)}]{Rajagopal96}
{Rajagopal}, M., \& {Romani}, R.~W. 1996, \apj, 461, 327

\bibitem[{{Ransom}(2002)}]{Ransom02}
{Ransom}, S.~M. 2002, in Astronomical Society of the Pacific Conference Series,
  Vol. 271, Neutron Stars in Supernova Remnants, ed. {P.~O.~Slane \&
  B.~M.~Gaensler}, 361--+

\bibitem[{{Ransom} {et~al.}(2003){Ransom}, {Cordes}, \&
  {Eikenberry}}]{Ransom03}
{Ransom}, S.~M., {Cordes}, J.~M., \& {Eikenberry}, S.~S. 2003, \apj, 589, 911

\bibitem[{{Ransom} {et~al.}(2001){Ransom}, {Greenhill}, {Herrnstein},
  {Manchester}, {Camilo}, {Eikenberry}, \& {Lyne}}]{Ransom01}
{Ransom}, S.~M., {Greenhill}, L.~J., {Herrnstein}, J.~R., {et~al.} 2001, \apjl,
  546, L25

\bibitem[{{Ransom} {et~al.}(2014){Ransom}, {Stairs}, {Archibald}, {Hessels},
  {Kaplan}, {van Kerkwijk}, {Boyles}, {Deller}, {Chatterjee},
  {Schechtman-Rook}, {Berndsen}, {Lynch}, {Lorimer}, {Karako-Argaman}, {Kaspi},
  {Kondratiev}, {McLaughlin}, {van Leeuwen}, {Rosen}, {Roberts}, \&
  {Stovall}}]{Ransom14}
{Ransom}, S.~M., {Stairs}, I.~H., {Archibald}, A.~M., {et~al.} 2014, \nat, 505,
  520

\bibitem[{{Romani}(1987)}]{Romani87}
{Romani}, R.~W. 1987, \apj, 313, 718

\bibitem[{{Rutledge} {et~al.}(2002){Rutledge}, {Bildsten}, {Brown}, {Pavlov},
  \& {Zavlin}}]{Rutledge02a}
{Rutledge}, R.~E., {Bildsten}, L., {Brown}, E.~F., {Pavlov}, G.~G., \&
  {Zavlin}, V.~E. 2002, \apj, 578, 405

\bibitem[{{Rutledge} {et~al.}(2004){Rutledge}, {Fox}, {Kulkarni}, {Jacoby},
  {Cognard}, {Backer}, \& {Murray}}]{Rutledge04}
{Rutledge}, R.~E., {Fox}, D.~W., {Kulkarni}, S.~R., {et~al.} 2004, \apj, 613,
  522

\bibitem[{{Servillat} {et~al.}(2012){Servillat}, {Heinke}, {Ho}, {Grindlay},
  {Hong}, {van den Berg}, \& {Bogdanov}}]{Servillat12}
{Servillat}, M., {Heinke}, C.~O., {Ho}, W.~C.~G., {et~al.} 2012, \mnras, 423,
  1556

\bibitem[{{Steiner} {et~al.}(2010){Steiner}, {Lattimer}, \&
  {Brown}}]{Steiner10}
{Steiner}, A.~W., {Lattimer}, J.~M., \& {Brown}, E.~F. 2010, \apj, 722, 33

\bibitem[{{Steiner} {et~al.}(2016){Steiner}, {Lattimer}, \&
  {Brown}}]{Steiner16}
---. 2016, European Physical Journal A, 52, 18

\bibitem[{{Suleimanov} {et~al.}(2011){Suleimanov}, {Poutanen}, \&
  {Werner}}]{Suleimanov11}
{Suleimanov}, V., {Poutanen}, J., \& {Werner}, K. 2011, \aap, 527, A139

\bibitem[{{Sztajno} {et~al.}(1987){Sztajno}, {Fujimoto}, {van Paradijs},
  {Vacca}, {Lewin}, {Penninx}, \& {Trumper}}]{Sztajno87}
{Sztajno}, M., {Fujimoto}, M.~Y., {van Paradijs}, J., {et~al.} 1987, \mnras,
  226, 39

\bibitem[{{van Paradijs}(1979)}]{Paradijs79}
{van Paradijs}, J. 1979, \apj, 234, 609

\bibitem[{{Vaughan} {et~al.}(1994){Vaughan}, {van der Klis}, {Wood}, {Norris},
  {Hertz}, {Michelson}, {van Paradijs}, {Lewin}, {Mitsuda}, \&
  {Penninx}}]{Vaughan94}
{Vaughan}, B.~A., {van der Klis}, M., {Wood}, K.~S., {et~al.} 1994, \apj, 435,
  362

\bibitem[{{Verbunt} \& {Hasinger}(1998)}]{Verbunt98}
{Verbunt}, F., \& {Hasinger}, G. 1998, \aap, 336, 895

\bibitem[{{Webb} \& {Barret}(2007)}]{Webb07}
{Webb}, N.~A., \& {Barret}, D. 2007, \apj, 671, 727

\bibitem[{{Wilms} {et~al.}(2000){Wilms}, {Allen}, \& {McCray}}]{Wilms00}
{Wilms}, J., {Allen}, A., \& {McCray}, R. 2000, \apj, 542, 914

\bibitem[{{Woodley} {et~al.}(2012){Woodley}, {Goldsbury}, {Kalirai}, {Richer},
  {Tremblay}, {Anderson}, {Bergeron}, {Dotter}, {Esteves}, {Fahlman}, {Hansen},
  {Heyl}, {Hurley}, {Rich}, {Shara}, \& {Stetson}}]{Woodley12}
{Woodley}, K.~A., {Goldsbury}, R., {Kalirai}, J.~S., {et~al.} 2012, \aj, 143,
  50

\bibitem[{{Worpel} {et~al.}(2013){Worpel}, {Galloway}, \& {Price}}]{Worpel13}
{Worpel}, H., {Galloway}, D.~K., \& {Price}, D.~J. 2013, \apj, 772, 94

\bibitem[{{Zamfir} {et~al.}(2012){Zamfir}, {Cumming}, \& {Galloway}}]{Zamfir12}
{Zamfir}, M., {Cumming}, A., \& {Galloway}, D.~K. 2012, \apj, 749, 69

\bibitem[{{Zavlin} \& {Pavlov}(2002)}]{Zavlin02b}
{Zavlin}, V.~E., \& {Pavlov}, G.~G. 2002, in Neutron Stars, Pulsars, and
  Supernova Remnants, ed. W.~{Becker}, H.~{Lesch}, \& J.~{Tr{\"u}mper}, 263

\bibitem[{{Zavlin} {et~al.}(2000){Zavlin}, {Pavlov}, {Sanwal}, \&
  {Tr{\"u}mper}}]{Zavlin00}
{Zavlin}, V.~E., {Pavlov}, G.~G., {Sanwal}, D., \& {Tr{\"u}mper}, J. 2000,
  \apjl, 540, L25

\bibitem[{{Zavlin} {et~al.}(1996){Zavlin}, {Pavlov}, \& {Shibanov}}]{Zavlin96}
{Zavlin}, V.~E., {Pavlov}, G.~G., \& {Shibanov}, Y.~A. 1996, \aap, 315, 141

\end{thebibliography}

\end{document}